\documentclass[10pt, conference, letterpaper]{IEEEtran}
\usepackage{amssymb}        
\usepackage[lined,boxed,linesnumbered,ruled,vlined]{algorithm2e}
\SetNlSty{}{}{:}	        
\SetKwInOut{Input}{Input}  
\SetKwInOut{Output}{Output}
\SetKwInOut{Returns}{Return}  
\SetKwInOut{Initialize}{Initialize} 
\usepackage[noend]{algorithmic} 
\usepackage{pgf}                
\usepackage{tabularx}   
\usepackage{url}        
\usepackage{amssymb} 

\usepackage[lined,boxed,linesnumbered,ruled,vlined]{algorithm2e} 
\SetKwInOut{Input}{Input}  
\SetKwInOut{Output}{Output}	
\SetKwInOut{Returns}{Return}  
\SetKwInOut{Initialize}{Initialize} 
\SetKwComment{Comment}{$ \%\ $}{}
\usepackage[noend]{algorithmic} 

\usepackage[autostyle]{csquotes}

\usepackage{amsmath}

\usepackage{subcaption}
\usepackage{tablefootnote}

\usepackage{dblfloatfix} 

\title{Shrewd Selection Speeds Surfing: Use Smart EXP3!}
\author{\IEEEauthorblockN{Anuja Meetoo Appavoo, Seth Gilbert, and Kian-Lee Tan}
\IEEEauthorblockA{Department of Computer Science, National University of Singapore}
\{anuja, seth.gilbert, tankl\}@comp.nus.edu.sg}
\begin{document}

\maketitle
\begin{abstract} 
In this paper, we explore the use of multi-armed bandit online learning techniques to solve distributed resource selection problems. As an example, we focus on the problem of network selection. Mobile devices often have several wireless networks at their disposal. While choosing the right network is vital for good performance, a decentralized solution remains a challenge.
%
The impressive theoretical properties of multi-armed bandit algorithms, like EXP3, suggest that it should work well for this type of problem. Yet, its real-word performance lags far behind. The main reasons are the hidden cost of switching networks and its slow rate of convergence. 
We propose Smart EXP3, a novel bandit-style algorithm that (a) retains the good theoretical properties of EXP3,
(b) bounds the number of switches, and (c) yields significantly better performance in practice.
We evaluate Smart EXP3 using simulations, controlled experiments, and 
in-the-wild
experiments. Results show that it stabilizes at the optimal state, 
achieves fairness among devices and gracefully deals with transient behaviors. In real world experiments, it can achieve 
18\% faster download over alternate strategies.
%
We conclude that multi-armed bandit algorithms can play an important role in distributed resource selection problems, when practical concerns, such as switching costs and convergence time, are addressed.

\end{abstract}

\section{Introduction} \label{section:introduction}                  
    \newcommand{\para}[1]{\noindent \textbf{#1:}~~}
\let\svthefootnote\thefootnote
\let\thefootnote\relax\footnote{This research was supported in part by AcRF Tier 1 grant T1 251RES1719.}
\addtocounter{footnote}{-1}\let\thefootnote\svthefootnote
Mobile devices often have several wireless networks at their disposal. Choosing the right network is vital for good performance. Yet, it is non-trivial. This is, in part, because network availability is transient and the quality of networks changes dynamically due to mobility of devices and environmental factors.
The conventional wisdom is to choose WiFi over cellular, and to associate with a WiFi Access Point (AP) that has the highest signal strength---which is often suboptimal~\cite{biswas2015large}. The challenge is for each device to make decentralized decisions, without any coordination, and yet achieve a \emph{fair} allocation, where each device gets an equal share of the available bandwidth (to the extent that it is feasible). Given that the environment is dynamic, it is harder to achieve an optimal solution. 
Resource selection problems can be formulated as a congestion game. Multi-armed bandit problem relates to repeated multi-player games, where each player independently aims at 
improving its decision and all other players collectively act as an adversary.
Furthermore, theoretical properties of multi-armed bandit algorithms suggest that they provide an excellent solution to this problem. 

EXP3 (Exponential-weight algorithm for Exploration and Exploitation)~\cite{auer2002nonstochastic}, one of the leading bandit algorithms, is fully decentralized and Hannan-consistent, i.e., as time elapses, it performs nearly as well as always selecting the best action in hindsight. It has been proven to converge to a (weakly stable) Nash equilibrium~\cite{kleinberg2009multiplicative,tekin2011performance} while guaranteeing good performance (i.e., minimizing regret). However, we observe (via simulation) that EXP3 tends to perform worse than even simple naive greedy solutions. The main reasons for the unexpectedly poor outcomes are (a) EXP3 does not capture switching cost, which is a 
non-negligible
cost in network selection, and (b) it has a relatively slow convergence; in some of our simulations, it took the equivalent of over 14 days to stabilize. We do not want to treat switching cost as a \enquote{loss}, from the perspective of EXP3, as this will unfairly penalize networks with high data rates and high switching cost. Moreover, while the process of exploring networks is designed to minimize regret, it does not optimize for quick convergence to a Nash equilibrium. Both of these problems are exacerbated in dynamic wireless network settings.

We formulate the wireless network selection problem as a repeated congestion game (in each round, each device chooses a network and receives some reward, i.e., bandwidth), and model the behavior of devices using online learning in the adversarial bandit setting. 
We propose Smart EXP3, a novel bandit-style algorithm that retains the good properties of EXP3 while addressing the issues that prevent it from achieving good performance in practice. From a theoretical perspective, we focus on the static version of the problem; in our experiments, we explore dynamic settings. There are a few key insights underlying Smart EXP3. The first observation is that we can minimize the cost of switching networks by using \emph{adaptive blocking} techniques.
The second observation is that we can speed up the rate of reaching a \enquote{stable state}
by carefully adding \emph{initial exploration} and a \emph{greedy policy}.  The third observation is that once the system is stable, we want to remain in a good state; we rely on a \emph{switch-back} mechanism.  Finally, in a dynamic setting, a careful \emph{minimal reset mechanism} is needed to ensure that the system adapts efficiently to changes.




To summarize, 
the following are our key contributions:
\begin{enumerate}
    \item We formulate the wireless network selection problem as a repeated congestion game and model the behavior of devices using online learning in bandit setting. 
    \item We show that EXP3 has relatively poor performance in a dynamic wireless network setting.
    \item We propose Smart EXP3, an algorithm that has good theoretical and practical performance.
    \item We demonstrate 
    (using simulations, controlled experiments, and in-the-wild experiments)
    that Smart EXP3 (a) gracefully deals with transient behaviors, 
    (b) stabilizes at the optimal state relatively fast, with reduced switching, and 
    (c) achieves fairness among devices. Since experiments 
    are more \enquote{expensive} to conduct than simulations, we compare the performance of Smart EXP3 to only that of the \enquote{best} performing alternative
    (from simulation results) in our experiments. However, we perform more extensive simulations.
    \item We give an upper bound on the expected number of network switches and prove that Smart EXP3 has the same convergence and regret properties as EXP3.
\end{enumerate}

A major goal of this paper is to discover how to make bandit-style algorithms (like EXP3) more effective in practice, without compromising on theoretical properties, by
focusing on important practical issues of switching cost, time to stabilize, and adaptation to transient behaviors.   
All source code for the simulations and real-world experiments, and data from real-world experiments are available 
on GitHub\footnote{https://github.com/anuja-meetoo/SmartEXP3}.
\section{Wireless network selection} \label{section:problemFormulation}
    In this section, we describe the wireless network selection problem, and formulate it as a repeated congestion game.
\subsection{Wireless network selection problem}
\begin{figure}[!htb]
\begin{center}
 \includegraphics[width=75mm,trim=134 0 110 0 mm, clip=true]{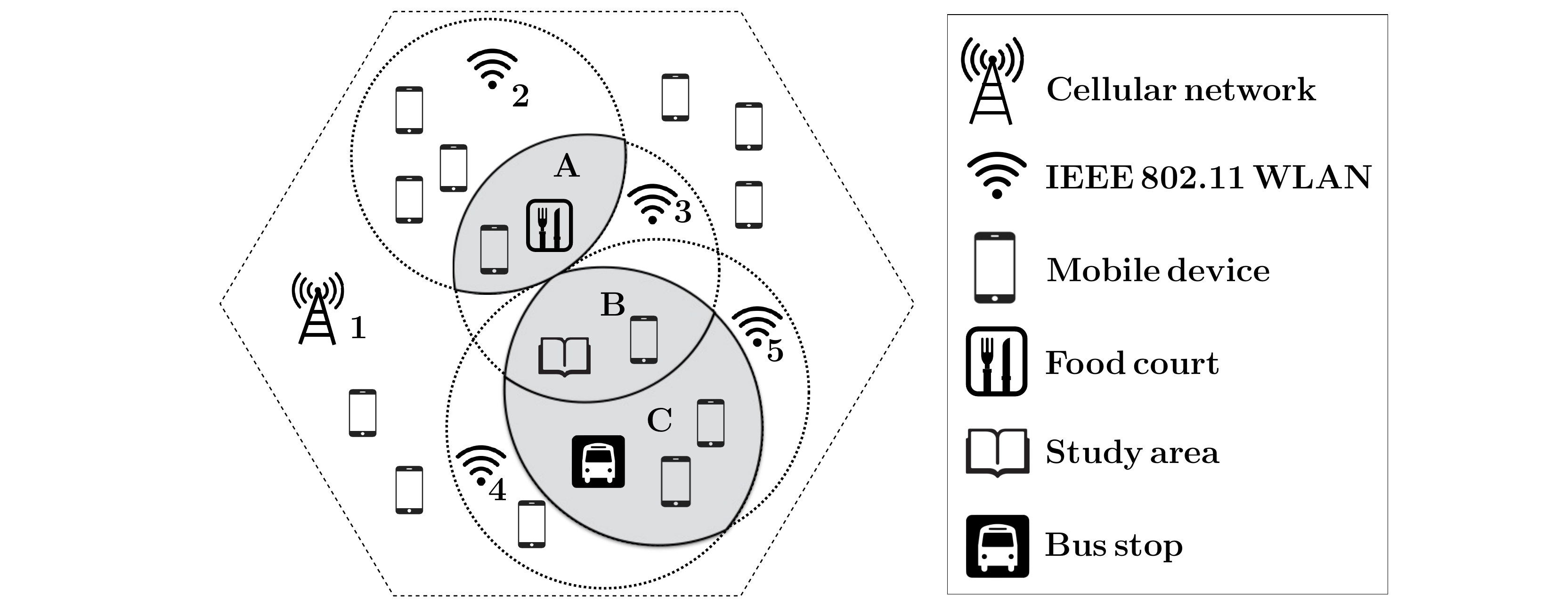} 
\end{center}
\caption{Service areas with heterogeneous wireless networks.}
\label{figure:heterogeneousNetworks}
\end{figure}

We consider a collection of mobile devices operating in an environment with heterogeneous networks. For example, Figure~\ref{figure:heterogeneousNetworks} depicts mobile devices operating in three service areas (shaded areas A, B and C) with several wireless networks.
The wireless networks are numbered from 1 to 5 and the dotted lines delimit their coverage. Different devices have access to different networks, e.g., devices at the food court will see the cellular network and WLANs 2 and 3. The goal is to connect each device to the best network, which may vary over time.
%
%

Three criteria are important when selecting a network: (a) the quality of the connection, which is influenced by the distance between a device and the AP, or the level of external interference; (b) the bandwidth of the network; and (c) the level of congestion, e.g., the number of devices sharing the network. 
While this information is not available to a device at the time of selection, the achievable data rates can be estimated by exploring the networks. 
Every time a device switches network, it incurs a cost, which we assume is measured in terms of delay, and sacrifices some available bandwidth. 
\subsection{Formulation of wireless network selection game}
\label{section:game_fomulation}
Since mobile devices operate in a dynamic environment, continuous exploration and adaptation are required. Wireless network selection 
can be formulated as a repeated resource selection game, a special type of congestion game \cite{rosenthal1973class}. 

We formally define the wireless network selection game as a tuple 
$\Gamma = \langle \mathcal{N}, \mathcal{K}, (\mathcal{S}_j)_{j \in \mathcal{N}}, (\mathcal{U}^t_i)_{i \in \mathcal{K}}\rangle$, where
\begin{enumerate}
\item $\mathcal{N} = \{1 \cdots n\}$ is the finite set of \textit{n} active mobile devices indexed by \textit{j}.
\item $\mathcal{K} = \{1 \cdots k\}$ denotes the finite set of \textit{k} wireless networks available 
in the service area. 
\item $\mathcal{S}_j \subseteq 2^{\mathcal{K}_j}$ is the strategy set of mobile device $j$, where $\mathcal{K}_j \subseteq \mathcal{K}$ is the set of networks available to $j$.
\item Gain (payoff or utility) $g_{i_j}(t)$ of mobile device $j$ refers to the bit rate it observes when selecting network $i_j$ at time $t$, scaled to [0, 1]; it is expressed by a function $\mathcal{U}_{i_j}$ of the number of devices $n_{i_j}(t)$ associated with $i_j$ as follows:
$$n_{i_j}(t) = |\{j' \in \mathcal{N}:i_{j'} = i_j\}| $$
where $i_{j'}$ is the network selected by $j'$ at time $t$.
$$ g_{i_j}(t) = \mathcal{U}_{i_j}^t(n_{i_j}(t)) $$ 

A device's gain affects its strategy and, hence, ignores switching cost so that networks with high gain but high switching cost are not penalized.
\item Cumulative goodput of a device $j$ 
is given by $$\sum_{t=1}^{T} \mathcal{U}_{x_j}^t(n_{x_j}(t)) \cdot (slot\_duration - delay)$$
where $delay$ is the switching cost ($delay$ is zero when the device stays in the same network), $slot\_duration$ (higher than $delay$) is the length of a time slot (assuming time is slotted), and $T$ is the time horizon.
\item A strategy profile is given by $\mathcal{S} = \mathcal{S}_1$ x $\cdots$ x $\mathcal{S}_n$. It is at Nash equilibrium \cite{nisan2007algorithmic} if $g_{i_j}(\mathcal{S}) \ge g_{i_j}(\mathcal{S}_{-j}, \mathcal{S}_j')$ for every $\mathcal{S}_j'$ and every $j \in \mathcal{N}$, where $(\mathcal{S}_{-j}, \mathcal{S}_j')$ implies that only device $j$ changes its strategy. Hence, no device wants to unilaterally change its strategy.
\end{enumerate}


The wireless network selection problem is related to the adversarial bandit problem \cite{auer2002nonstochastic}, in which a gambler must select a slot machine to play in a sequence of trials to maximize the cumulative reward. In our case, the aim of each device $j$ is to maximize its cumulative 
goodput
by quickly identifying and connecting to the best network. The performance of a network degrades proportionally to the number of devices supported; other mobile devices accessing shared networks may be considered adversaries. We model the behavior of devices using online learning in the adversarial bandit setting, where EXP3 \cite{auer2002nonstochastic} is a standard algorithmic solution. 
Each device performs an independent network selection and the
only information available to 
it is its set of available networks.


\section{Smart EXP3} \label{section:proposedAlgorithm}
In this section, we develop Smart EXP3, a distributed wireless network selection algorithm, by diligently modifying EXP3 \cite{auer2002nonstochastic} so as to retain its good properties while compensating for its shortcomings. 
It runs independently on each mobile device. Yet, it affects the choice of other devices that have a common set of available networks (by affecting their gains). 

\noindent\textbf{EXP3.} We briefly explain how EXP3 \cite{auer2002nonstochastic} works. It maintains a weight for each network, which represents the confidence that the network is a good choice.
Initially, a device assumes uniform weight over all networks. The weight of a network is affected by the gain (bit rate) the device observes by associating with it; a higher gain implies higher weight. EXP3 assumes that time is slotted. At each time slot, it selects a network randomly from a probability distribution, that mixes between using the weights and a uniform distribution; the latter ensures that EXP3 keeps exploring occasionally and discovers a better network that was previously \enquote{bad}. The best network will eventually gain higher weight and be selected most often.

\noindent\textbf{Differences of Smart EXP3 compared to EXP3.}
There are three major differences. First, it selects a network for a longer duration of time, using \emph{adaptive blocking}. Second, it has an initial exploration phase and occasionally leverages a \emph{greedy policy} to make a deterministic selection, while EXP3 always performs a random selection. Third, it allows a device to switch back to its previous network upon selecting a worse network. 

\noindent\textbf{
Adaptive blocking.}
Each device partitions time into \emph{blocks}, and selects a network to associate with for the entire block. Each block consists of a sequence of time slots of equal length. The duration of a time slot is long enough for a device to observe the gain.
%
The block length used by a device grows over time and is given by $\lceil(1 + \beta)^x\rceil$, where $\beta \in (0, 1]$ and $x$ is the number of times the network has been selected by that device. This ensures that more time is spent in the optimal network, which is eventually selected more frequently. The use of blocks reduces switching cost \cite{auer2002finite, du2016learning,chen2011Opportunistic} and improves performance by de-synchronizing the selection time of devices. Every so often and upon significant decline in network quality, block lengths are reset for better adaptation. 



\noindent\textbf{Algorithm description.} Algorithm \ref{algorithm:smartEXP3} outlines the major steps in Smart EXP3, excluding the parts on reset and updates made when a change in the set of available networks is detected. 
See Table~\ref{table:variables} for notations. 
%
We defer explanation on switch back. 

\noindent Much like EXP3, Smart EXP3 assigns a weight to each network. At the beginning of a block, the probability distribution is updated based on the weights of the networks. The same multiplicative weight update and probability update rules as for EXP3 \cite{auer2002nonstochastic} are used. Smart EXP3 then selects a network $i_b$ to associate with during the whole block. In the first $k$ blocks, it explores the networks in random order, and $\overline{p}(b) = \frac{1}{|explore\_network|}$. This improves the learning rate. From block $k + 1$ onward, it either selects randomly based on its probability distribution or considers the use of a greedy approach. In the prior case, $\overline{p}(b) = p_{i_b}(b)$. 
The mobile device observes a gain during the entire block, which is used to update the network's weight at the end of the block.
%
The estimated gain $\hat{g_{i_b}}(b)$ in the weight update rule compensates for a potentially small probability of observing the gain.

\renewcommand\thempfootnote{\arabic{mpfootnote}}
\begin{table}[!htb]
\small
\centering
\caption{Notations used to describe Smart EXP3.}
\label{table:variables}
\newcolumntype{L}{>{\arraybackslash}m{5.5cm}}
\begin{tabular}{c L}
\hline
$\mathcal{K}$ & Set of networks available.                        \\ 
$k$                  & $|\mathcal{K}|$                           \\ 
$explore\_network$   & Set of networks not yet explored.                 \\ 
$w_i$                & Confidence that network $i$ is a good choice.     \\ 
$p_i$                & Probability for choosing network $i$.             \\ 
$i_b$                & Network chosen for block $b$.                     \\ 
$\overline{p}$     & Probability with which $i_b$ was chosen. \tablefootnote{It depends on the type of selection made, i.e., whether it was an initial exploration, a random choice, a greedy selection, or a switch back.}             \\ 
$l_i$                & Block length of network $i$.                      \\ 
$g_i(b) \in [0,l_{i}]$   & Gain observed from network $i$ in block $b$.       \\ 
$x_i$                & No. of blocks in which network $i$ is chosen.     \\ 
\hline
\end{tabular}
\end{table}
\begin{algorithm}[!htb]
  \DontPrintSemicolon 
  \Input{$k \in \mathbb{Z}_{>0}$, real $\gamma \in (0, 1] $, real $\beta \in (0, 1] $}
  \Initialize{$ w_{i}(1) \leftarrow 1 $ for $ i = 1, \cdots, k $, \\ $explore\_network \leftarrow \mathcal{K}$}
  \BlankLine 
  \ForEach{block $b = 1, 2, \cdots $}{
    
    $ p_{i}(b) \leftarrow (1 - \gamma)\frac{w_{i}(b)}{\sum\limits_{j=1}^{k} w_{j}(b)} + \frac{\gamma}{k} $ for $i = 1, \cdots , k$\\
    
    \If{explore\_network $\neq \O$}
    {$i_{b} \leftarrow $ random from $explore\_network$ \\
    $explore\_network \leftarrow explore\_network\ \backslash\ \{i_b\}$}
    \ElseIf{chooseGreedily() = True} {$i_{b} \leftarrow $ network with highest average gain}
    \lElse{$i_{b} \leftarrow $ random according to $p(b)$}
    
    $l_{i_b} = \lceil(1 + \beta)^{{x}_{{i}_{b}}}\rceil$
    \BlankLine
    \Comment{execute block for $l_{i_b}$ time slots.}
    \Comment{at the second time slot, switch back to the previous network if the current one is worse, and start a new block.}
    \BlankLine
    
    $g_{i_{b}}(b) \leftarrow $ gain observed, where $g_{i_{b}}(b) \in [0, l_{i_b}]$\\

    $\hat{g_{i_{b}}}(b) \leftarrow \frac{g_{i_{b}}(b)}{\overline{p}(b)}$
    
    $ w_{i_{b}}(b + 1) \leftarrow w_{i_{b}}(b)\ exp(\frac{\gamma \hat{g_{i_{b}}}(b)}{k})$\\
  }
  
  \caption{Smart EXP3 \newline
  \textit{Shows the major steps in the algorithm, leaving out the parts on (1) reset, and (2) updates made when a change in the set of available networks is detected.
  \newline
  chooseGreedily() determines whether \enquote{greedy} selection can be leveraged; a device selects greedily with probability $\frac{1}{2}$ at the beginning of an excecution, or for some time after a reset.
  }}
  \label{algorithm:smartEXP3}
\end{algorithm}

\noindent\textbf{Greedy choices.} At the beginning of an execution, or for some time after a reset, the mobile device flips an unbiased coin and decides (with equal probability) to use either a greedy or a random strategy. In the prior case, it selects the seemingly \enquote{best} network, i.e., the network from which the highest average gain has been observed. Then, $\overline{p}(b) = \frac{1}{2}$. If the device decides to choose randomly, $\overline{p}(b) = \frac{p_{i_b}(b)}{2}$.
An aggressive use of greedy selection generally leads to low efficiency in social welfare.
However, allowing half the devices to choose greedily, at first, causes them to perturb the weight of their perceived \enquote{best} network and allows other devices to explore and adapt. Empirical results show that it drastically improves the rate at which the algorithm stabilizes. 


\noindent\textbf{Switching back.} If a device switches network when the algorithm is at Nash equilibrium, it will observe a lower gain. Based on this intuition, if a device observes a worse performance during the first time slot of a block, it starts a \emph{special} block at the next time slot. In that block, the mobile device simply associates to its previous network rather than executing lines 3 - 8 of Algorithm \ref{algorithm:smartEXP3}. Here, $\overline{p}(b) = 1$. Smart EXP3 does not allow a device to switch back in two consecutive blocks to prevent a ping-pong effect. The switch back mechanism reduces the time spent in a bad network (restricts it to a block of a single time slot), and prevents other devices from reacting. Empirical results show that this mechanism makes Smart EXP3 much more stable. 


\noindent\textbf{Minimal reset.} 
Smart EXP3 must converge to the optimal network, and yet quickly respond to changes in the environment. For instance, when the probability of one particular network is sufficiently high causing the device to stay in that network for a long time, the algorithm becomes less adaptive to changes. It might take an unacceptable amount of time to discover resources freed by other devices. Hence, Smart EXP3 resets every so often, and when it detects a significant drop in quality of the network being selected for consecutive time slots. At that point, network block lengths and details stored for use during greedy selection are reset.
It then forces exploration of available networks. As such, reset is minimal to allow the algorithm to adapt without forsaking everything it has learned. The duration between two resets is referred to as a reset period.

\noindent\textbf{Change in set of networks.} When a device discovers a new network, its weight is set to the maximum weight of the other networks or 1 if all networks are newly discovered; then, the algorithm resets. In addition, the algorithm resets when a network with significantly high probability of being selected is no longer available.
These ensure that a newly discovered network is likely to be explored and the algorithm adapts quickly to the change. If the network to which the device was connected is no longer available, Smart EXP3 resets the block.

\section{Theoretical analysis of Smart EXP3} \label{section:formalAnalysis}
    \newtheorem{theorem}{Theorem}
Due to the changes we have made to EXP3, it is not immediately apparent that Smart EXP3 has the same convergence and regret properties as EXP3.
We show that it does and give an upper bound on its number of switches. 

The duration of a time slot is denoted by $t_d$ and a reset period by $\tau$.
For the purpose of the analysis, we assume that (a)
$\mathcal{K}_i = \mathcal{K}$
for every $j \in \mathcal{N}$, i.e., all devices have the same set of networks available to it, and (b) the environment is static.

\noindent\textbf{Convergence.} 
Strategies in the support of the mixed strategy $\delta_j$ of player $j$ are those played with a non-zero probability \cite{nisan2007algorithmic}. Weakly stable equilibria \cite{kleinberg2009multiplicative} is defined as mixed Nash equilibria $(\delta_1, \cdots, \delta_n)$ with the additional property that each player $j$ remains indifferent between the strategies in the support of $\delta_j$ when any other single player $j'$ changes to a pure strategy in the support of $\delta_{j'}$; however, each strategy in the support of $\delta_j$ may not remain a best response and device $j$ may prefer a strategy outside the support of $\delta_j$.


We consider Smart EXP3 without reset and prove that it retains the convergence property of EXP3. 
We show that the dynamics of the probability distribution over the set of available networks is given by a replicator equation which is identical to the one of EXP3 \cite{tekin2011performance}.

\begin{theorem}
When $\gamma$ is arbitrarily small, the strategy profile of all devices using Smart EXP3 converges to a weakly stable 
equilibrium; weakly stable 
equilibria are pure Nash equilibria with probability $1$ when the bit rate of each network is chosen at random independently \cite{kleinberg2009multiplicative}.
\end{theorem}
Hence, when all devices leverage Smart EXP3, they end up being optimally distributed across networks. No device will observe higher gain by unilaterally switching network. 
%
%
Although, it is not conveyed by the analysis, empirical results show that Smart EXP3 reaches a \emph{stable state} 
(defined in section \ref{section:simulationConvergence}) 3.3x faster than EXP3 in some settings considered.

The formal proof is provided in appendix \ref{appendix:convergenceToNE}.

\noindent\textbf{Bound on number of network switches.}
We 
bound the number of network switches. 
\begin{theorem}
\label{theorem:numNetworkSwitch}
For any $k > 0$, $\beta \in (0, 1]$, time slot duration $t_d \in  \mathbb{Z}_{\ge0}$, reset period $\tau > 0$, 
and stopping time $T > 0$,
the expected number of network switches over time $T$ is upper bounded as:
$$E[S(T)] < \frac{T}{\tau}\left(\frac{3\ k \ \log(\frac{\tau}{t_d} + 1)}{\log(1 + \beta)} \right) $$
\end{theorem}
The logarithmic bound implies that the rate at which Smart EXP3 switches networks decreases over time.

Assuming $t_d = 1$ and $\tau = T$ (i.e., there is no reset), $$E[S(T)] < \frac{3\ k \ \log(T + 1)}{\log(1 + \beta)}$$
It implies that longer time horizon $T$, and higher number of wireless networks (to explore) increase the number of switches. Faster growth of block size (controlled by $\beta$) will reduce the number of switches.

\noindent Referring to Theorem \ref{theorem:numNetworkSwitch}, we also infer that a higher delay (switching cost) implies longer time slots, and hence, reduced number of switches. Longer reset periods will also reduce the number of switches, as the latter decreases over time in a reset period. Empirical results show a drastic reduction in the number of network switches compared to that of EXP3.

The formal proof is given in appendix \ref{appendix:networkSwitchBound}.

\noindent\textbf{Regret bounds.} We define \emph{weak} regret as follows:
\newtheorem{definition}{Definition}
\begin{definition}
{\emph{Weak regret.}} It refers to the difference between the cumulative goodput (capturing switching cost) achieved by always selecting the best network in hindsight and that of Smart EXP3.
\end{definition}
We follow the proof of EXP3 \cite{auer2002nonstochastic} and 
show that
Smart EXP3 retains the logarithmic weak regret property of EXP3. 
Let $G_{SmartEXP3}(T)$ denote the cumulative gain of Smart EXP3 at T, $G_{max}(T)$ be the cumulative gain at T when always choosing the best network in hindsight, $\mu_d$ be the mean delay observed, and $\mu_g$ denote mean gain (bit rate) observed.
\begin{theorem}
\label{theorem:regretBound}
For any $k > 0$, any fixed $\gamma \in (0, 1]$, any $\beta \in (0, 1]$, any assignment of rewards, stopping time $T > 0$, time slot duration $t_d \in  \mathbb{Z}_{\ge0}$, reset period $\tau > 0$, the highest block length $l$, mean delay $\mu_d \ge 0$, and mean gain $\mu_g \ge 0$, the expected weak regret is upper bounded as:
$$E[R(T)] \le \frac{T \cdot t_d}{\tau}\left((1 + \gamma \ l\ (e - 2))\ G_{max}(\tau) + \frac{k\ \ln\ k}{\gamma}\right) $$
$$+ \frac{T \cdot \mu_d \cdot \mu_g}{\tau}\left(\frac{3\ k \ \log(\frac{\tau}{t_d} + 1)}{\log(1 + \beta)}\right) $$
%
\end{theorem}


\noindent Hence, Smart EXP3 is Hannan-consistent as its weak regret tends to zero. As time elapses, it performs nearly as well as always selecting the best network in hindsight.

Assuming $t_d = 1$ and $\tau = T$ (i.e., there is no reset), 
$$E[R(T)] \le (1 + \gamma \ l\ (e - 2))\ G_{max}(T) + \frac{k\ \ln\ k}{\gamma} $$
$$ + \mu_d \cdot \mu_g\left(\frac{3\ k \ \log(T + 1)}{\log(1 + \beta)}\right) $$
The first term implies that: (a) if the cumulative goodput achieved by always choosing the best network is high, the regret can be high (if the goodput of Smart EXP3 is low); in that case, having long blocks, increases the regret (which would imply Smart EXP3 is staying in a bad network for a long duration; but this is not seen in our evaluations), and (b) weak regret grows with an increase in number of networks (exploring sub-optimal networks). The second term implies that weak regret increases with a rise in (a) number of network switches, (b) mean delay observed, and (c) mean bit rate observed.

\noindent Referring to Theorem \ref{theorem:regretBound}, we also infer that long time slot duration yields an increase in regret,
as more time is spent in sub-optimal networks. Longer reset periods will reduce regret as the latter decreases over time in a reset period.



The formal proof is provided in appendix \ref{appendix:regretBound}.
\section{Implementation details} \label{section:implementationDetails}
    We thoroughly evaluate Smart EXP3 and compare its performance against those of several other algorithms, through simulation and experiments. All algorithms are implemented in Python, using SimPy \cite{simpy} for simulation. In this section, we discuss the implementation of Smart EXP3 focusing on the greedy, switch back and reset mechanisms. We discuss the parameter values chosen for the simulation and experiments. $p$ denotes the probability distribution, $i_+$ refers to the network with the highest probability, and $i_{max}$ denotes the network selected for the highest number of time slots.

\noindent\textbf{Parameter choice.} In our implementation, $\gamma = b^{\frac{-1}{3}}$\cite{maghsudi2013relay}, where $\gamma$ is the exploration rate and tends to zero to ensure convergence \cite{tekin2011performance}, and $b$ is the block index;
$\beta = 0.1$ such that blocks are short during exploration; and the duration of one time slot is 15 seconds (simulated seconds for simulation), i.e., greater than the maximum delay observed while switching networks during some experiments in real-world settings. 

\noindent\textbf{Greedy choices.}
Smart EXP3 considers the use of greedy when: (a) $max(p)-min(p)\le\frac{1}{k-1}$, given that it starts with a uniform probability, or (b) $l_{i_+} < y$, where $y$ is the value of $l_{i_+}$ when condition (a) evaluates to \emph{false} for the first time. The second condition allows for the use of greedy after a reset. Based on empirical results, these are good choices. When either of these conditions evaluates to \emph{true}, the device selects greedily with probability $\frac{1}{2}$ (flipping an unbiased coin).  

\noindent\textbf{Switch back.}
A device switches back if (a) the gain from the current network is worse than the average gain observed in the preceding block or during its last time slot, or if more than 50\% of the time, a higher gain was observed in the preceding block, and (b) the algorithm did not switch back at the beginning of the current block (to prevent a ping-pong effect). To ignore stale data, the decision is based on observations from only the last 8 time slots of the previous block. 

\noindent\textbf{Resetting.}
The algorithm resets when $p_{i_+} \ge 0.75$ and $l_{i_+} \ge 40$, i.e., the algorithm stays for a long duration in the network which has a sufficiently high probability of being selected. This allows for discovery of resources that have recently been freed. It also resets if a drop of at least 15\% is observed in $i_{max}$ to which the device is connected since more than 4 time slots. 
This ensures that the algorithm reacts to an actual change in the environment, rather than to noise with less than 15\% change or a change observed only during one time slot. 
These allow fast adaptation while preventing frequent resets.
\section{Evaluation through simulation} \label{section:evaluationSimulation}
    This section shows that EXP3 has poor performance in a dynamic wireless network setting. It then evaluates Smart EXP3, relying on simulations using synthetic data (Section \ref{section:evaluationSimulation}), and trace-driven simulations (Section \ref{section:evaluationSimulationNetworkTrace}).


\subsection{Simulation using synthetic data} \label{section:evaluationSimulation}
In this section, we show that EXP3 incurs high switching cost, has slow convergence, and fails to adapt to changes in the environment. In contrast, Smart EXP3 (a) stabilizes at Nash equilibrium with reduced switching, (b) better utilizes available resources, (c) achieves fairness among devices, (d) scales with an increase in number of devices and networks, (e) adapts to changes in the environment, and (f) is robust against \enquote{greedy} devices.
It outperforms alternative selection algorithms given in Table~\ref{table:algorithm_comparison}.
As baselines, we include a \emph{Full information} and a \emph{Centralized} protocol even though they can not be implemented without coordination among devices or via a base station, as they assume the availability of global knowledge.
The performance of algorithms in Table~\ref{table:algorithm_incremental_update} is also discussed to highlight benefits of key features of Smart EXP3.

\begin{table}[!htb]
\small
\centering
\caption{Algorithms to which Smart EXP3 is compared.}
\label{table:algorithm_comparison}
\newcolumntype{L}{>{\arraybackslash}m{5.5cm}}
\begin{tabular}{c L}
\hline
Full Information & It assigns a weight to each network. At each time slot, it selects a network at random based on their weights. At the end of a time slot, the device receives feedback about the gain it could obtain from each network, and computes the loss of each of them. The weight of each network is updated based on their loss, using a multiplicative update rule \cite{gyorgy2006adaptive}.
\\
\hline
Greedy & It explores each network in random order. Then, at each time slot, it selects a network 
with  highest average gain.
\\
\hline
Centralized & It is optimal (maintains Nash equilibrium) and assumes that a centralized entity allocates devices to the right network.
\\
\hline
Fixed Random & It picks a network at random and stays in that network. 
\\
\hline
\end{tabular}
\end{table}
\begin{table}[!htb]
\small
\centering
\caption{Algorithms highlighting features of Smart EXP3.}
\label{table:algorithm_incremental_update}
\newcolumntype{L}{>{\arraybackslash}m{5cm}}
\begin{tabular}{c L}
\hline
Block EXP3 & Version of EXP3 that selects a network for a block of time slots.
\\
\hline
Hybrid \newline Block EXP3 & Version of Block EXP3 which includes the greedy policy of Smart EXP3.
\\
\hline
Smart EXP3\newline w/o Reset & Smart EXP3 version that never resets.
\\
\hline
\end{tabular}
\end{table}

\noindent\textbf{Setup.} We consider two settings of 20 devices and 3 networks, with an aggregate bandwidth of 33 Mbps. Setting 1 assumes non-uniform data rates 4, 7 and 22 Mbps, a factor close to the theoretical data rates of IEEE 802.11 standards \cite{ieee80211} and cellular networks \cite{gessner2012umts} that yields a unique Nash equilibrium. In Setting 2, the networks have a uniform data rate (11 Mbps each). Delay is modeled using Johnson's SU distribution for WiFi and Student's t-distribution for cellular, each identified as a best fit~\cite{fitter} to 500 delay values.
We make the following assumptions (which are not pre-requirements for the algorithm) in the simulation: (1) a network's bandwidth is equally shared among its clients, and (2) clients are time-synchronized. \color{black} 
Results involve data from 500 runs of 5 (simulated) hours each, i.e., 1200 time slots, unless specified otherwise.

\begin{figure}[!htb]
\begin{center}
\includegraphics  [scale=0.75]
{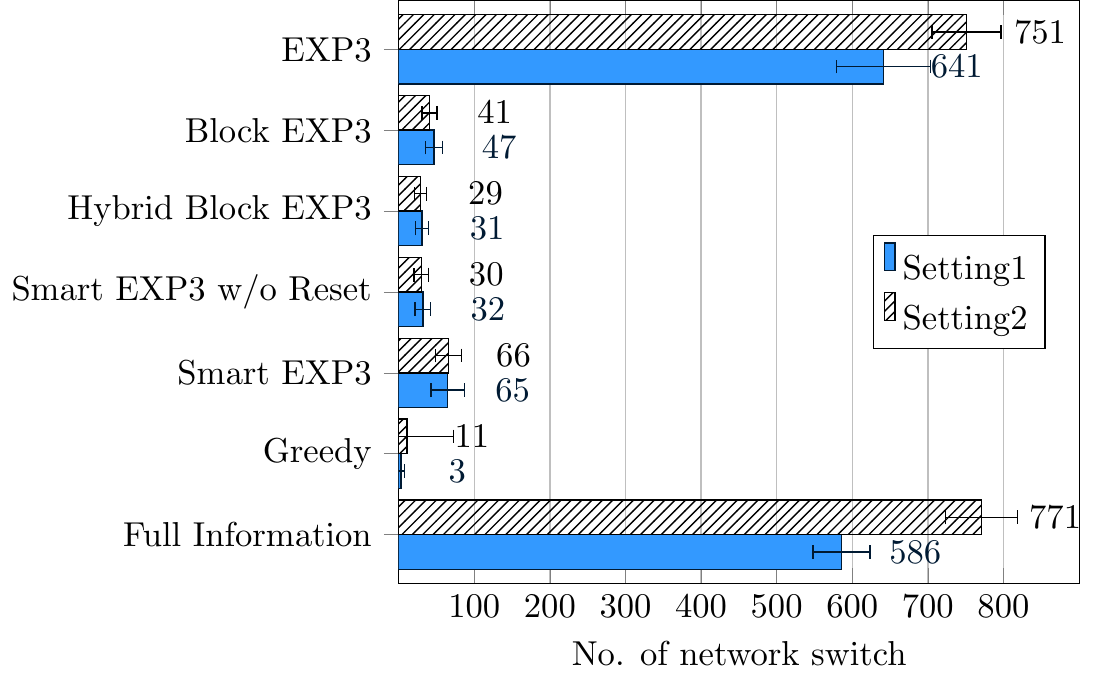}
\end{center}
\caption{Average number of network switches incurred by each algorithm in both settings (error bar shows standard deviation).}
\label{figure:static_numNetworkSwitch}
\end{figure}

\noindent\textbf{Switching cost.}
Figure \ref{figure:static_numNetworkSwitch} shows that EXP3 and Full Information incur high number of network switches.
Block-based algorithms experience around 80\% lower switching cost, and lower variance, in both settings. The costs of Hybrid Block EXP3 and Smart EXP3 w/o Reset are lower than that of Block EXP3 as their greedy policy helps them become stable faster, as discussed later. Thus, block lengths increase faster. The cost of Smart EXP3 increases with resets, but is acceptable. As discussed later, reset promotes faster adaptation to changes in network conditions. 
Greedy may incur high cost in setting 2, where 8 devices switched networks more than 83.3\% of time. Centralized and Fixed Random approaches 
do not incur switching cost, hence are not shown in Figure \ref{figure:static_numNetworkSwitch}.

\noindent\textbf{Stability and distance to Nash equilibrium.} \label{section:simulationConvergence}
We define the notion of 
\emph{stable state} to evaluate the algorithms' performance.
\begin{definition}{\emph{Stable state.}} An algorithm is said to have reached a stable state when each device selects a particular network with sufficiently high probability (we assume $\ge 0.75$), and maintains a sufficiently high probability for that same network until the end.
\label{def:stable_state}
\end{definition}

\begin{figure}[!htb]
\begin{center}
\includegraphics  [scale=0.8,trim=0 3 0 0 mm, clip=true]
{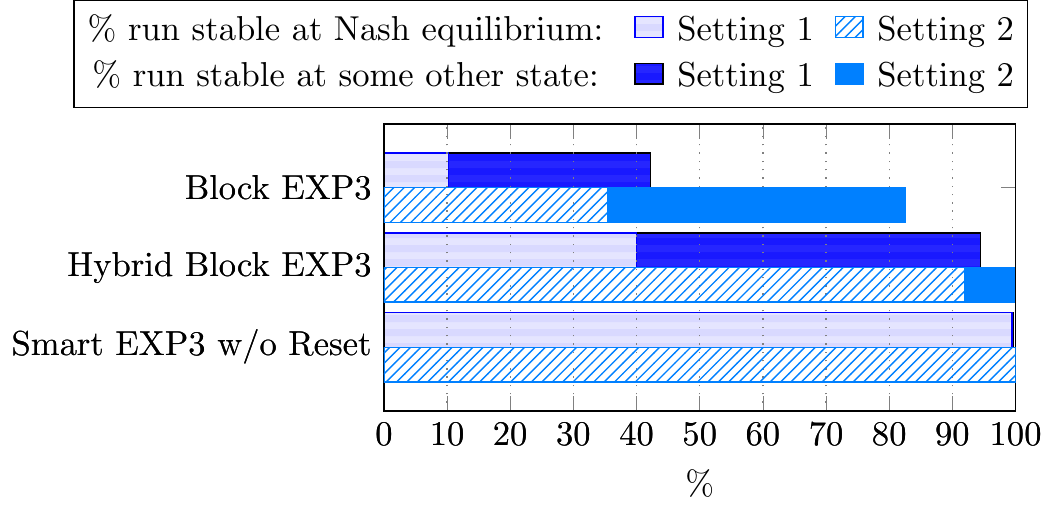}
\end{center}
\caption{Percentage run that reached a stable state and type of stable state (Nash equilibrium or some other state).}
\label{figure:static_perRunConvergence}
\end{figure}

EXP3 and Full Information never reached a stable state in our simulation, due to frequent switching. Figure \ref{figure:static_perRunConvergence} shows that more than 40\% of Block EXP3 runs stabilize, but rarely at Nash equilibrium. As given in Table~\ref{table:static_rateOfConvergence}, it takes very long to reach the stable state. The greedy policy in Hybrid Block EXP3 significantly improves the rate at which the algorithm stabilizes. The switch back mechanism retains Smart EXP3 w/o Reset in the optimal state, leading to $99.4\%$ and $100\%$ runs being stable at Nash equilibrium in settings 1 and 2, respectively, at a faster rate. As setting 2 has three Nash equilibria with an equal distribution of devices over networks, the algorithms perform better (their initial distribution is uniform). 
\begin{table}[!htb]
\small
\centering
\caption{Median no. of time slots taken to reach a stable state (whether Nash equilibrium or some other state).}
\label{table:static_rateOfConvergence}
\newcolumntype{L}{>{\centering\arraybackslash}p{2cm}}
\begin{tabular}{l L L L}
\hline
\textbf{}                              & \textbf{Block EXP3} & \textbf{Hybrid Block EXP3} & \textbf{Smart EXP3 w/o Reset} \\ \hline
\multicolumn{1}{l}{\textbf{Setting 1}} & 1026       & 583.5      & 359        \\ 
\multicolumn{1}{l}{\textbf{Setting 2}} & 810        & 366        & 244.5      \\ 
\hline
\end{tabular}
\end{table}

%


\begin{figure*}[!ht]
  \begin{subfigure}{1\columnwidth}
    \includegraphics [scale=1.2,trim=0 5 0 3 mm, clip=true,height=6.25cm]{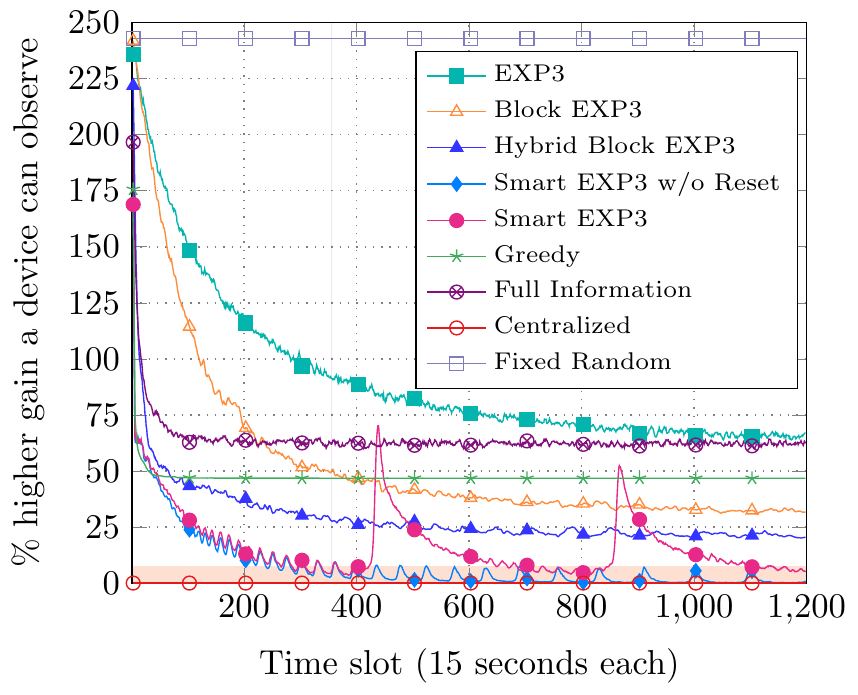}
    \caption{Setting 1.\\}
    \label{figure:static_4_7_22_20users_3networks_distanceToNE}
  \end{subfigure}
  \hfill 
  \begin{subfigure}[!hb]{1\columnwidth}
    \includegraphics[scale=1.2,trim=0 5 0 2 mm, clip=true,height=6.25cm]{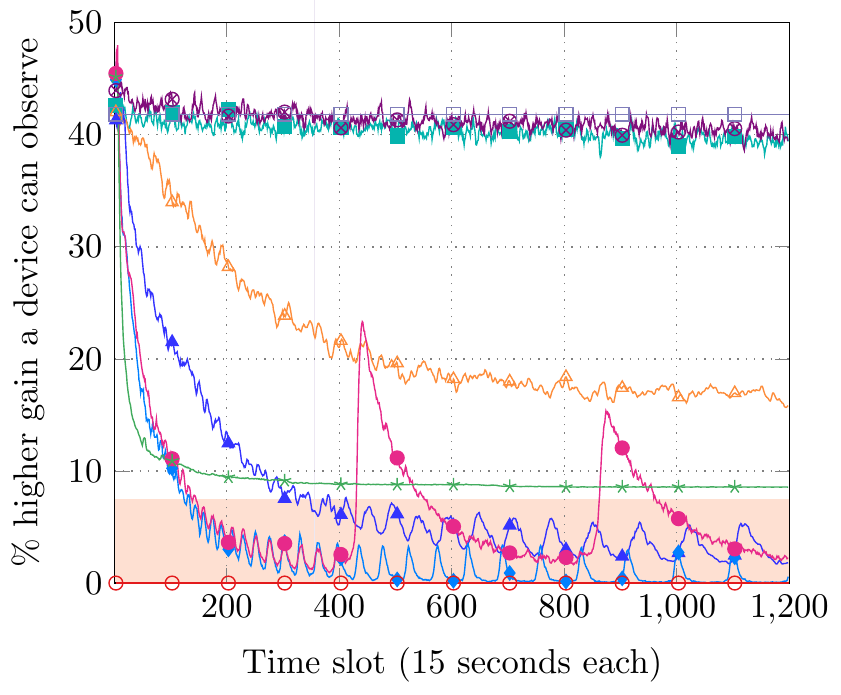}
    \caption{Setting 2 (legend is the same as that of Figure \ref{figure:static_4_7_22_20users_3networks_distanceToNE}).}
    \label{figure:static_11_11_11_20users_3networks_distanceToNE}
  \end{subfigure}
  \caption{Average distance to Nash equilibrium (\% higher gain any device would have observed, compared to its current gain, if the algorithm was at Nash equilibrium) --- shaded region represents $\epsilon$-equilibrium, where $\epsilon = 7.5$; in Figure \ref{figure:static_11_11_11_20users_3networks_distanceToNE}, EXP3, Full information and Fixed Random maintain a distance close to 40\%).}
  \label{figure:static_20users_3networks_distanceToNE}
\end{figure*}

Some algorithms cannot, by definition, be evaluated based on the notion of stable state, e.g., Greedy, Centralized, Fixed Random, and Smart EXP3 (due to resets). Thus, we define \emph{distance to Nash equilibrium} as a common evaluation criterion, to evaluate how the \emph{state} of an algorithm evolves over time (illustrated in Figure \ref{figure:static_20users_3networks_distanceToNE}).
The \emph{state} of an algorithm refers to the allocation of devices to networks, i.e., the number of devices associated to each network.

Strategy profile $\mathcal{S} = \mathcal{S}_1$ x $\cdots$ x $\mathcal{S}_n$ is at $\epsilon-equilibrium$ \cite{nisan2007algorithmic} if $g_{i_j}(\mathcal{S}) \ge g_{i_j}(\mathcal{S}_{-j}, \mathcal{S}_j') - \epsilon$ for every $\mathcal{S}_j'$ and every $j \in \mathcal{N}$, where $i_j$ is the network selected by device $j$, $g_{i_j}(\mathcal{S})$ is the gain observed by device $j$ given strategy profile $\mathcal{S}$, $\epsilon$ is a real non-negative parameter and $(\mathcal{S}_{-j}, \mathcal{S}_j')$ implies that only device $j$ changes its strategy. This implies that no device can achieve more than $\epsilon$ increase in gain by unilaterally deviating from its strategy. In line with the definition of $\epsilon-equilibrium$, we define the notion of \emph{distance to Nash equilibrium}. 

\begin{definition}{\emph{Distance to Nash equilibrium.}}
The distance between the current state of an algorithm and Nash equilibrium is given by the maximum percentage higher gain any device would have observed if the algorithm was at Nash equilibrium, compared to its current gain.
\end{definition}

As an example, we consider the setting with three mobile devices and two wireless networks. Assume that the three devices observe bit rates 1 Mbps, 1 Mbps and 4 Mbps. At Nash equilibrium, they would each observe 2 Mbps. Compared to their current gains, two devices would observe $100\%$ higher bit rate while the third one would observe a lower bit rate. The distance to Nash equilibrium is then considered to be $100\%$.


Figure \ref{figure:static_4_7_22_20users_3networks_distanceToNE} confirms that Smart EXP3 w/o Reset stabilizes at Nash equilibrium. As expected, Fixed Random performs badly. Greedy is stable, but at a \enquote{bad} state. Smart EXP3 outperforms all these algorithms, though its distance from equilibrium rises during periodic resets, as seen by the two spikes. While the reset mechanism does not seem useful in this static setting, it is vital for fast adaptation in a dynamic setting, as we shall see later. It also occasionally drifts away from the optimal state, shown as fluctuations, but is forced to return by the switch back mechanism.  It spends 62.77\% and 74.30\% time at Nash equilibrium in settings 1 and 2, respectively, and is at $\epsilon$-equilibrium most of the time, when $\epsilon = 7.5$. Figure \ref{figure:static_11_11_11_20users_3networks_distanceToNE} shows that distances in setting 2 are lower, as expected.




\noindent\textbf{Unutilized resources.}
In each setting, with aggregate bandwidth of 33 Mbps, the total bandwidth available over 1200 time slots (15 seconds each) is 74.25 GB. As Greedy starts by exploring available networks in a random order, it is highly likely that $\frac{1}{3}$ of the devices will be associated with each network during exploration. Hence, in setting 1, most devices are likely to rate the network with 4 Mbps bandwidth as unusable and end up selecting one of the other two networks. This leads to a situation similar to \enquote{tragedy of the commons}, with unutilized resources. It loses 8 GB on average in that setting, but utilizes all resources in setting 2. The other algorithms ensure that devices discover and utilize all resources on average in both settings, although not always in an optimal or fair way. 

\noindent\textbf{Cumulative download and fairness.}
The number of network switches and state at which an algorithm stabilizes affect its cumulative goodput. Table
~\ref{table:simulation_cumulative_median_gain} shows that the  block-based algorithms achieve higher cumulative goodput, on average. Greedy has lower performance than Smart EXP3 in setting 1 but comparable performance in setting 2, as expected. Fixed Random also achieves comparable performance in setting 2. 

\begin{figure}[!htb]
\begin{center}
\includegraphics  [scale=0.35,trim=14 265 22 7 mm, clip=true] 
{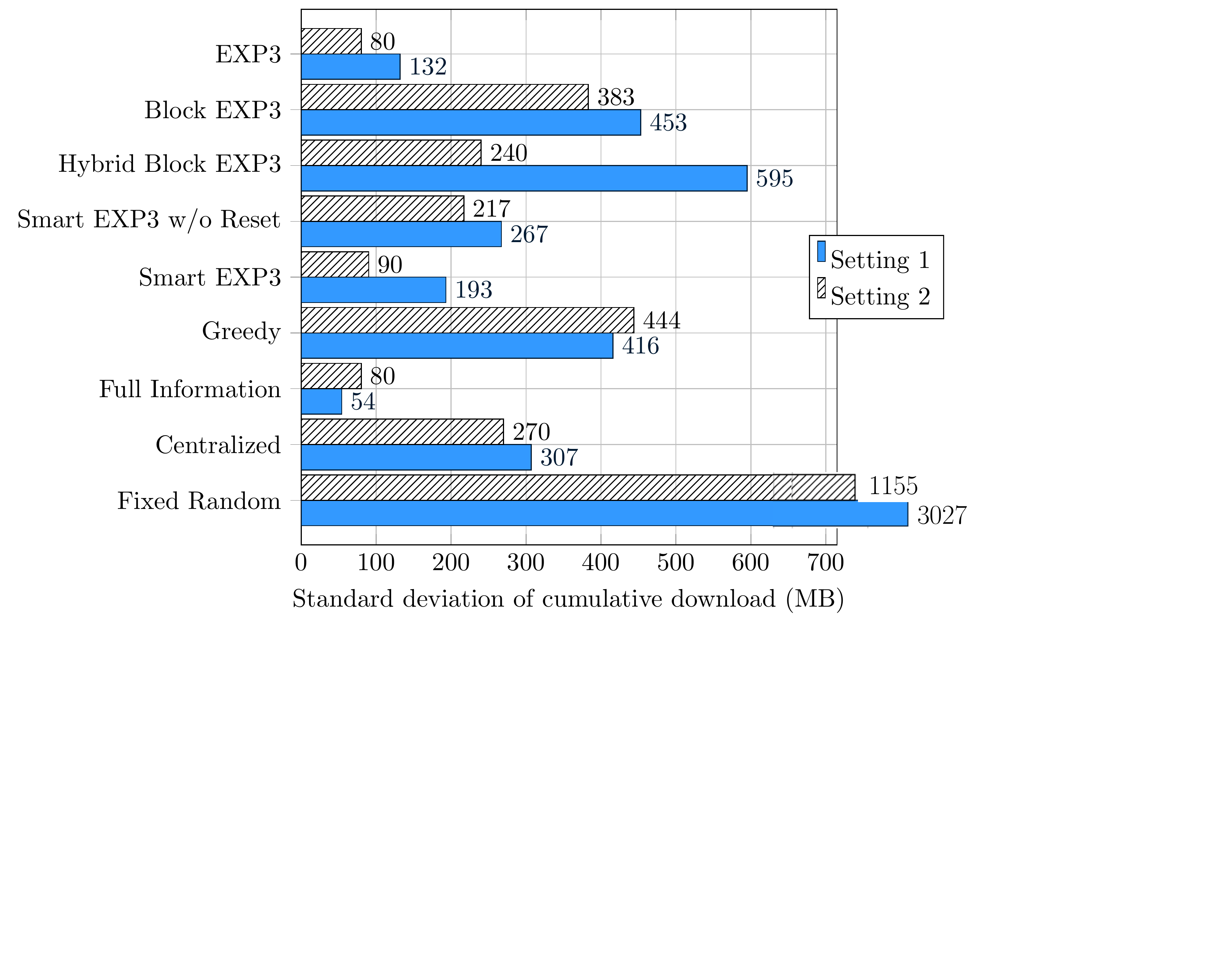}
\end{center}
\caption{Average per run standard deviation of cumulative download (MB) achieved by devices.}
\label{figure:static_20users_3networks_cummulativeDownload_average_per_run_std}
\end{figure}

We evaluate fairness of an algorithm in terms of the standard deviation of the cumulative downloads of individual devices. A lower standard deviation implies a fairer allocation where more devices achieve about the same cumulative download. On the other hand, a higher standard deviation means more disparity among individual cumulative downloads.
Figure \ref{figure:static_20users_3networks_cummulativeDownload_average_per_run_std} shows that EXP3, Smart EXP3 and Full Information are fairer among the algorithms. The standard deviations of Smart EXP3 are 80\% and 55\% less than those of Greedy in settings 1 and 2, respectively. While Nash equilibrium may not be fair, periodic reset can lead to fairness if devices converge to a different network after a reset. Although, Smart EXP3 switches more often than Greedy, it is worth spending time exploring to achieve higher and fairer cumulative download.


\begin{table}[!htb]
\small
\caption{(Mean) per run median cumulative download (GB).}
\label{table:simulation_cumulative_median_gain}
\centering 
\begin{tabular}{lcc}
\hline
& \multicolumn{2}{c}{\centering\textbf{Median cumulative download (GB)}} \\
& \textbf{Setting 1}  & \textbf{Setting 2}   \\ \hline
\multicolumn{1}{l}{EXP3} & 2.89  & 2.73 \\ 
\multicolumn{1}{l}{Block EXP3} & 3.54  & 3.65 \\ 
\multicolumn{1}{l}{Hybrid Block EXP3} & 3.41 & 3.58 \\ 
\multicolumn{1}{l}{Smart EXP3 w/o Reset} & 3.53 & 3.55 \\ 
\multicolumn{1}{l}{Smart EXP3} & 3.53 & 3.62 \\ 
\multicolumn{1}{l}{Greedy} & 3.12 & 3.62 \\ 
\multicolumn{1}{l}{Full Information} & 2.92 & 2.71 \\ 
\multicolumn{1}{l}{Centralized} & 3.54 & 3.54 \\ 
\multicolumn{1}{l}{Fixed Random} & 2.56 & 3.43 \\ 
\hline
\end{tabular}
\end{table} 

\begin{figure}[!htb]
\begin{center}
\includegraphics  [scale=0.68]
{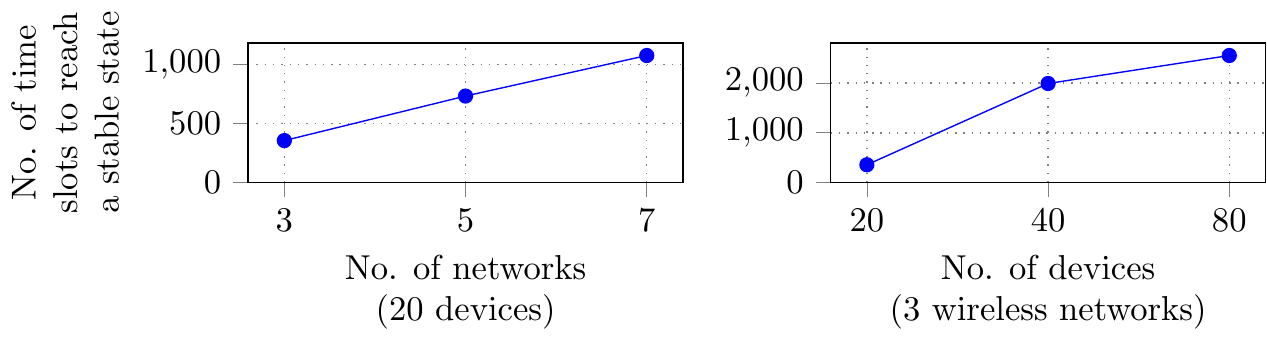}
\end{center}
\caption{Mean number of time slots taken by Smart EXP3 w/o Reset to reach a stable state, with increase in number of networks and devices.}
\label{figure:static_scalability}
\end{figure}

\noindent\textbf{Scalability.}
Scalability is evaluated in terms of the rate at which an algorithm reaches a \emph{stable state} (Definition \ref{def:stable_state}). Since Smart EXP3 cannot be evaluated based on this concept, Smart EXP3 w/o Reset is considered here. The algorithm was run 500 times, for 8640 time slots (i.e., 36 simulated hours) each, with different number of devices and networks. Figure \ref{figure:static_scalability} shows the median number of time slots taken to stabilize, in each setting. The rate increases linearly with an increase in number of networks and sub-linearly with an increase in number of devices. Furthermore, Smart EXP3 w/o Reset was stable at Nash equilibrium 100\% (or nearly 100\%) of times in each of the settings considered.

\noindent\textbf{Adaptability to changes in the environment.}
So far, we have seen that Greedy performs better compared to Full Information and Fixed Random. Thus, we only evaluate the performance of EXP3, Smart EXP3, Smart EXP3 w/o Reset and Greedy in 3 dynamic settings, with 20 devices each. 

\begin{figure}[!htb]
\begin{center}
\includegraphics [scale=0.84]
{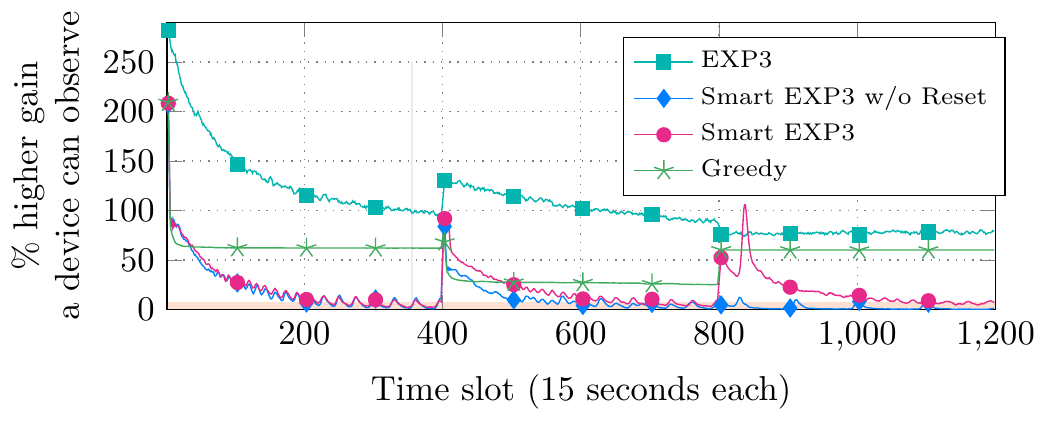}
\end{center}
\caption{Average distance to Nash equilibrium (\% higher gain any device would have observed, compared to its current gain, if the algorithm was at Nash equilibrium) --- shaded region represents $\epsilon$-equilibrium, where $\epsilon = 7.5$; 9 devices join at $t = 401$ and leave at the end of $t = 800$.}
\label{figure:dynamic_setting_1_20users_3networks_distanceToNE}
\end{figure}

\begin{figure}[!htb]
\begin{center}
\includegraphics [scale=0.84]
{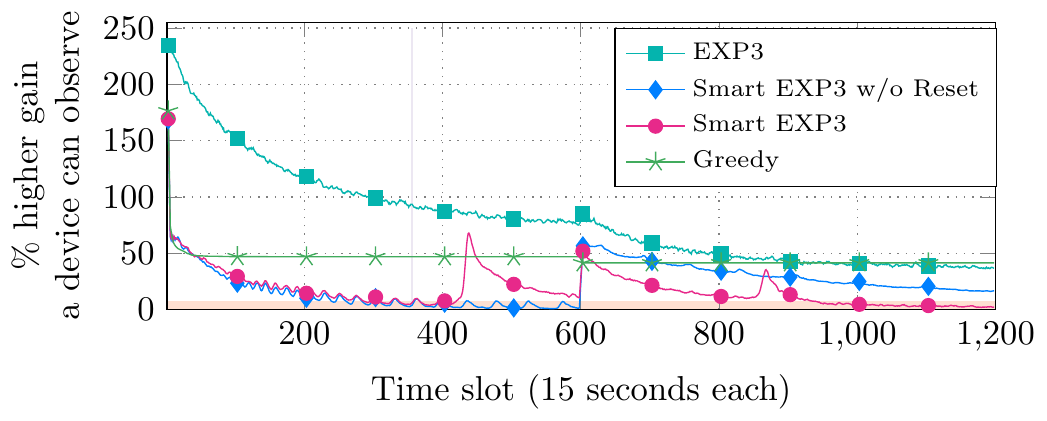}
\end{center}
\caption{Average distance to Nash equilibrium (\% higher gain any device would have observed, compared to its current gain, if the algorithm was at Nash equilibrium) --- shaded region represents $\epsilon$-equilibrium, where $\epsilon = 7.5$; 16 devices leave at the end of $t = 600$.}
\label{figure:dynamic_setting_2_20users_3networks_distanceToNE}
\end{figure}

In settings 1 and 2, all devices see 3 networks with bandwidth 4, 7 and 22 Mbps. In setting 1, 9 devices join at the beginning of $t = 401$ and leave at the end of $t = 800$, while the others are always in the service area. Figure \ref{figure:dynamic_setting_1_20users_3networks_distanceToNE} shows that only Smart EXP3 and Smart EXP3 w/o Reset are able to adapt to these changes. Their average distances to Nash equilibrium increase when the 9 devices join and begin exploring, but they eventually converge (at least very close) to the optimal allocation. In setting 2, 16 devices leave at the end of time slot $t = 600$, freeing resources. Figure \ref{figure:dynamic_setting_2_20users_3networks_distanceToNE} shows that only Smart EXP3 is able to discover the resources and adapt accordingly, highlighting the importance of the minimal reset mechanism.

\begin{figure}[!h]
\begin{center}
\includegraphics [scale=0.8]
{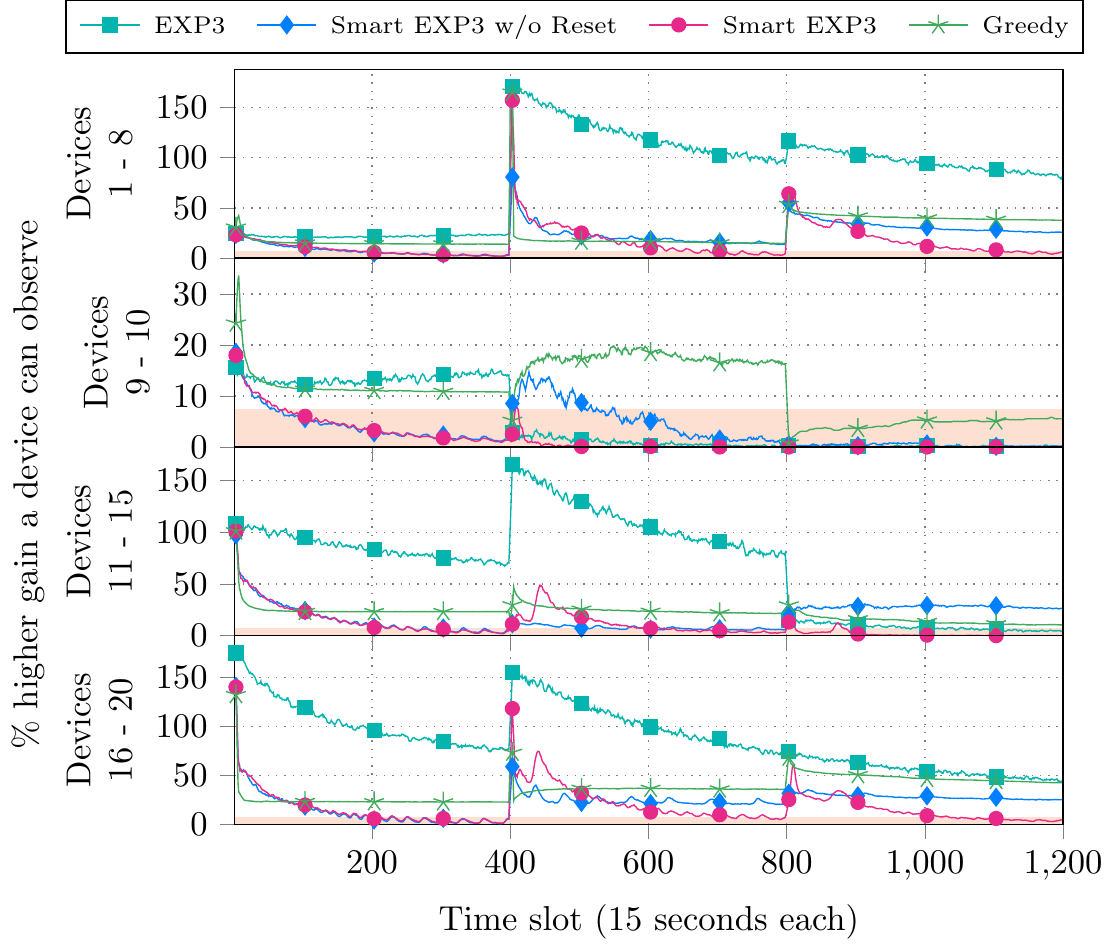}
\end{center}
\caption{Average distance to Nash equilibrium (\% higher gain any device would have observed, compared to its current gain, if the algorithm was at Nash equilibrium) --- shaded region represents $\epsilon$-equilibrium, where $\epsilon = 7.5$; 8 devices moving.}
\label{figure:mobility_setting_20users_3networks_distanceToNE}
\end{figure}

Setting 3 considers devices moving across service areas in Figure \ref{figure:heterogeneousNetworks}. Networks 1, 2, 3, 4 and 5 have bandwidth 16, 14, 22, 7 and 4, respectively. Initially there are 10 devices (1 -10) at the food court, 5 devices (11 - 15) at the study area and 5 devices (16 - 20) at the bus stop. 8 devices (1 - 8) from the food court move to the study area at the beginning of $t = 401$ and eventually reach the bus stop at the start of $t = 801$. Figure \ref{figure:mobility_setting_20users_3networks_distanceToNE} illustrates the performance of the algorithms for devices
in each area and those moving across areas, separately.
Smart EXP3 outperforms all the other algorithms for each category of devices and 
evolves
to at least $\epsilon-equilibrium$, when $\epsilon=7.5$.

\begin{figure}[!h]
\begin{center}
\includegraphics [scale=0.65]
{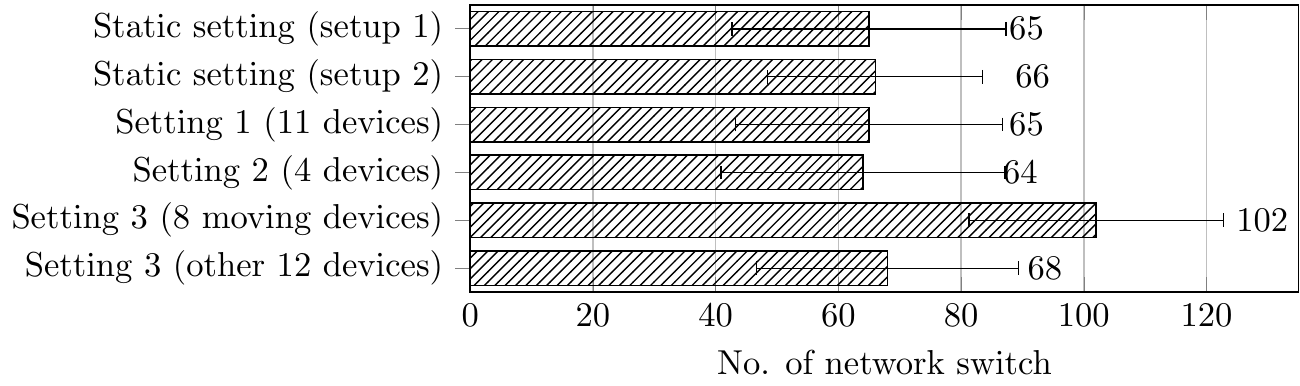}
\end{center}
\caption{Average number of network switches incurred by devices who stays in the experiment throughout (error bar shows standard deviation) --- in different settings.}
\label{figure:dynamic_setting_perClientNumNetworkSwitch}
\end{figure}

Figure \ref{figure:dynamic_setting_perClientNumNetworkSwitch} shows that the number of network switches incurred by devices in static and dynamic settings are comparable. Devices which are moving are likely to incur higher number of resets (median of 3 compared to median of 2 for temporarily stationary devices in our case), hence higher number of network switches. This is because Smart EXP3 resets when it discovers new networks and when a device's preferred network is no longer available.

\begin{figure}[!htb]
\begin{center}
\includegraphics  [scale=0.78]
{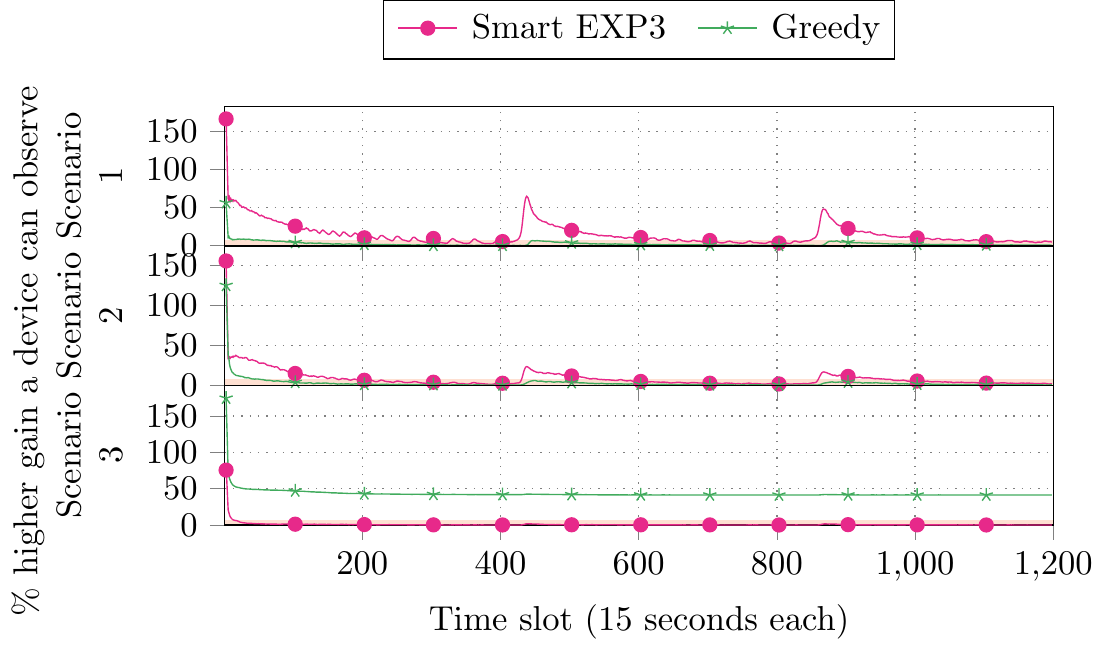}
\end{center}
\caption{Average distance to Nash equilibrium (\% higher gain any device would have observed, compared to its current gain, if the algorithm was at Nash equilibrium); settings with some devices using Smart EXP3 and others using Greedy.}
\label{figure:dishonestUsers}
\end{figure}

\noindent\textbf{Robustness against \enquote{greedy} devices.}
The performance of Smart EXP3 is evaluated in a setting, with 20 devices and 3 networks, where some devices use Greedy. In scenario 1, a single device uses Greedy while the others use Smart EXP3. In scenario 2, 10 devices employ each of the selection algorithms. And, in scenario 3 a single device uses Smart EXP3 while the others use Greedy. Figure \ref{figure:dishonestUsers} shows that, while Greedy is able to achieve good results in scenarios 1 and 2, it yields poor performance when the number of \enquote{greedy} users increase in scenario 3. On the other hand, Smart EXP3 performs well in all three scenarios and is robust against \enquote{greedy} devices.


\subsection{Trace-driven simulation} \label{section:evaluationSimulationNetworkTrace}
    Results from Section \ref{section:evaluationSimulation} show that Greedy performs better among alternative approaches.
Hence, we evaluate the performance of Smart EXP3, in comparison to that of Greedy only, based on network traces.
We collected traces of a public WiFi network and a cellular network by downloading a file from a remote server \cite{speedTest} on both networks simultaneously and measuring their bit rates. We evaluate the algorithms on 4 pairs of network traces, of 25 minutes each. The bit rates fluctuate, especially for the cellular network, although cellular network is always better than WiFi in trace 2. Results presented, from 500 simulation runs, show that Smart EXP3 adapts to changing network conditions and achieves higher cumulative goodput.


Table~\ref{table:trace_based_simulation_result} gives the cumulative download and switching cost incurred 
by each algorithm, when run on each of the 4 pairs of network traces. Smart EXP3 outperforms Greedy with traces 1, 3 and 4, where no single network is always the \enquote{optimal} choice. 
Greedy, however, performs well with trace 2 where the cellular network is always better. While Smart EXP3 explores WiFi from time to time, it spends most of the time in cellular network, achieving nearly the same performance as Greedy.
\begin{table}[!htb]
\small
\caption{Median of cumulative download (MB) and total switching cost (MB) incurred by Smart EXP3 and greedy.}
\label{table:trace_based_simulation_result}
\centering 
\newcolumntype{L}{>{\centering\arraybackslash}p{1.3cm}}
\begin{tabular}{l L L L L} 
\hline
                              & \multicolumn{2}{c}{\textbf{Smart EXP3}}        & \multicolumn{2}{c}{\textbf{Greedy}}       \\ 
                              & \textbf{Download}  & \textbf{Cost}    & \textbf{Download}  & \textbf{Cost}   \\ \hline
\multicolumn{1}{l}{Trace 1} & 764.16 & 39.74 & 671.07 & 3.05 \\ 
\multicolumn{1}{l}{Trace 2} & 1188.56 & 32.48 & 1235.92 & 6.14 \\ 
\multicolumn{1}{l}{Trace 3} & 657.81 & 44.11 & 428.47 & 2.96 \\ 
\multicolumn{1}{l}{Trace 4} & 810.67 & 51.11  & 757.66 & 4.50 \\ \hline
\end{tabular}

\end{table} 

\begin{figure}[!htb]
\begin{center}
\includegraphics  [scale=0.87]
{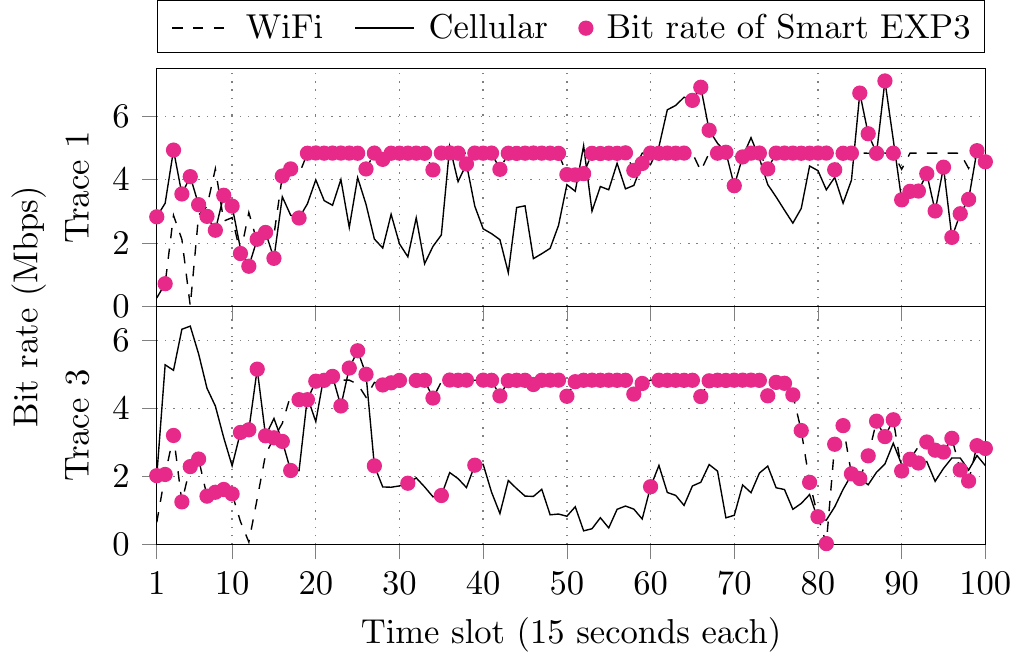}
\end{center}
\caption{Two pairs of simultaneous traces of a public WiFi and a cellular network, and an illustration of the network section process by Smart EXP3 at every time slot (shown as bit rate observed) - a random run with approximately the same cumulative download as the median cumulative download.}
\label{figure:trace_driven_simulation}
\end{figure}

Figure \ref{figure:trace_driven_simulation} illustrates the network selection process in one random run of Smart EXP3  (a run with cumulative download which is close to the median cumulative download) on traces 1 and 3, showing how it adapts to changes in network conditions.

\section{Evaluation through experiments in real-world settings} \label{section:evaluationExperiment}
    This section evaluates Smart EXP3 based on controlled experiments (Section \ref{section:evaluationExperimentControlled})
to see how it works in a real world setting where we still have some control over the network bandwidth and the number of devices, and in-the-wild experiments (Section \ref{section:evaluationExperimentWild}),
carried out in a coffee shop involving public networks. Since Greedy performs better among alternative approaches, we only compare the performance of Smart EXP3 to that of Greedy and show that it is better. 


\subsection{Controlled experiments}\label{section:evaluationExperimentControlled}
In this section, we show 
that Smart EXP3 outperforms Greedy, in terms of cumulative goodput achieved, efficiency of resource utilization and adaptability to changes in network conditions in real world settings. 

\noindent\textbf{Setup.} The setup consists of (a) 3 WiFi APs that operate on channels 11, 6 and 1 of the 2.4GHz band and with total bandwidth of 4, 7 and 22 Mbps;
(b) 2 laptops, each running a TCP server that continuously sends data to its clients (a request is sent to an alternate server when one fails to respond); (c) 14 raspberry pis that act as clients; and (d) a main AP that connects the servers and 3 WiFi APs through LAN cables. Devices run Smart EXP3 or Greedy and receive data from the server. They are synchronized, with a drift of less than one second. Switching networks is implemented by closing and establishing new wireless network and TCP connections. Gain is estimated based on the download during the time spent in a network. Results are based on 10 runs of 2 hours each, i.e. 480 time slots of 15 seconds.

\noindent\textbf{Switching cost, download and resource utilization.}
As expected, Smart EXP3 incurs a higher number of network switches (median of 73.5) compared to Greedy (median of 3). However, this enables the algorithm to explore and eventually achieve higher and fairer cumulative download, as shown in Table~\ref{table:controlled_experiment_cumulative_gain}.
Given the real world challenges, it also incurs a higher number of network switches (median of 73.5 in 2 hours compared to 61 in 5 simulated hours) and resets (median of 5 in 2 hours compared to 2 in 5 simulated hours) than in simulation. Furthermore, results show that it utilizes resources better than Greedy, which incurs a mean loss of 3.74\% of the aggregate resources. 

\begin{table}[!htb]
\small
\caption{Per run median cumulative download (as a \% of the estimated total download possible, based on bit rates observed by the devices).}
\label{table:controlled_experiment_cumulative_gain}
\centering 
\newcolumntype{L}{>{\centering\arraybackslash}m{2.4cm}}
\newcolumntype{M}{>{\centering\arraybackslash}m{2.6cm}}
\begin{tabular}[t]{l L M}
\hline
& \multicolumn{2}{c}{\textbf{Cumulative download (\%) of a single device}} \\
& \textbf{(Average) median}  & \textbf{(Average) standard deviation}   \\ \hline
\multicolumn{1}{l}{Smart EXP3} & 6.89 & 1.55 \\ 
\multicolumn{1}{l}{Greedy} & 6.29 & 2.87 \\ 
\hline
\end{tabular}
\end{table}




 


\noindent\textbf{Distance from average bit rate available.}
Network bit rates observed fluctuate due to factors such as interference and packet loss. In addition, a device may not observe an equal share of a network's bandwidth, e.g., due to its distance from the AP. 
As such, the notions of Nash equilibrium and stable state are hard to apply. Hence we define the notion of \emph{distance from average bit rate available}.
\begin{definition}{\emph{Distance from average bit rate available.}}
We estimate the bandwidth of each network based on bit rates observed by the devices, taking into account delay incurred when switching to the network. 
We calculate the average bandwidth $g$ available for each device as the aggregate bandwidth of all networks divided by the number of devices.
We then compute the average amount that observed bit rates fall below $g$, i.e., average of all $ max(g - g_{i_j}, 0) * \frac{100}{g}$, where $g_{i_j}$ is the bit rate observed by device $j$.
\end{definition}

\begin{figure}[!htb]
\begin{center}
\includegraphics  [scale=0.87]
{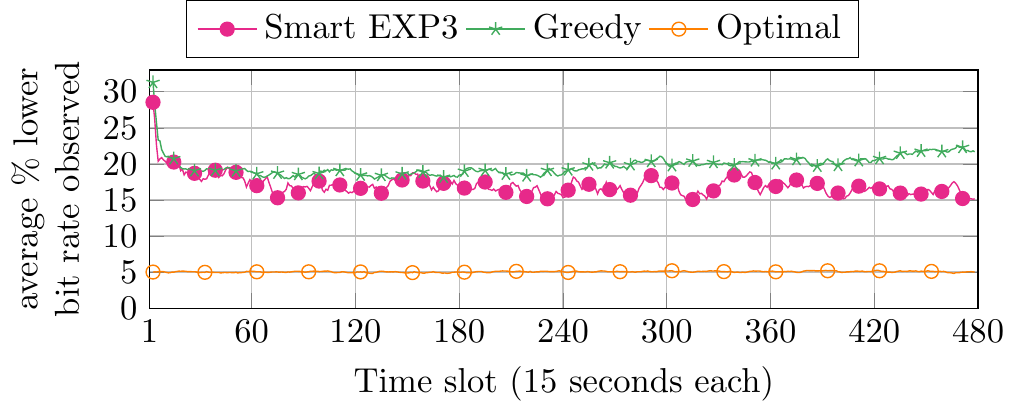}
\end{center}
\caption{Mean distance from average bit rate available in a static setting.}
\label{figure:static_4_7_22_14users_3networks_distanceFromAvgAvailable}
\end{figure}

\begin{figure}[!htb]
\begin{center}
\includegraphics  [scale=0.87]
{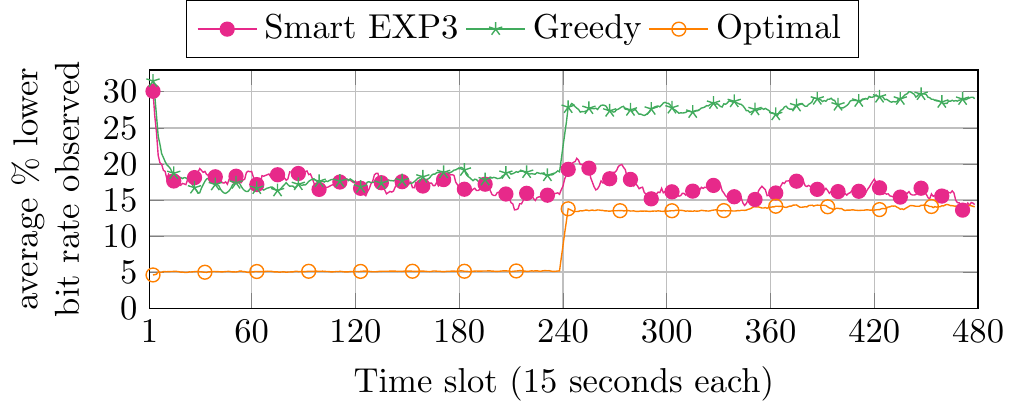}
\end{center}
\caption{Mean distance from average bit rate available in a dynamic setting --- 9 devices leave at the end of $t = 240$.}
\label{figure:dynamic_4_7_22_14users_3networks_distanceFromAvgAvailable}
\end{figure}

\begin{figure}[!htb]
\begin{center}
\includegraphics  [scale=0.87]
{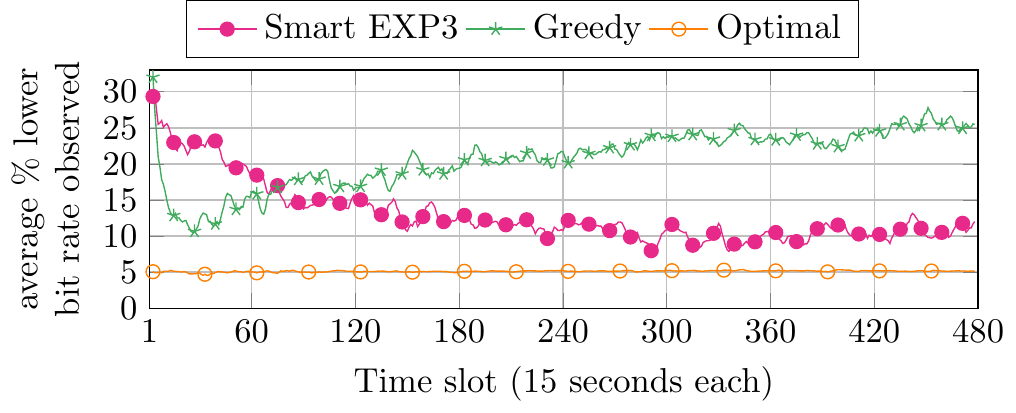}
\end{center}
\caption{Mean distance from average bit rate available when 50\% devices use Smart EXP3 and 50\% leverage Greedy.}
\label{figure:static_4_7_22_14users_3networks_dishonestUsers}
\end{figure}

The \emph{optimal} distance from the average bit rate available, shown in Figures
\ref{figure:static_4_7_22_14users_3networks_distanceFromAvgAvailable},
\ref{figure:dynamic_4_7_22_14users_3networks_distanceFromAvgAvailable}
and
\ref{figure:static_4_7_22_14users_3networks_dishonestUsers}
indicates the minimum distance that can be achieved at Nash equilibrium, given the estimated bandwidth of each network. We assume, here, that a network's bandwidth is equally shared among devices associated with it.

\noindent Figure \ref{figure:static_4_7_22_14users_3networks_distanceFromAvgAvailable} shows that the distance for Greedy 
gradually increases as the bit rates observed by some of the devices go down for some reason and the algorithm fails to adapt. The distance for Smart EXP3 eventually drops as the devices explore, learn and adapt, hence switching to a better network. However, noise in the real world perturbs the accuracy of the estimate of network quality and leads to a higher number of resets, preventing the distance from dropping any further.

\noindent\textbf{Adaptability to changes in the environment.}
A dynamic setting is considered in which 9 devices leave at the end of time slot $t = 240$, i.e. after 1 hour. Figure \ref{figure:dynamic_4_7_22_14users_3networks_distanceFromAvgAvailable} shows that both algorithms exhibit similar behaviors as in the static setting in the first 240 time slots.  When the devices leave at $t = 240$, resources are freed. The distance of Smart EXP3 rises at that time slot. But, given that Smart EXP3 continuously explores its environment, it is able to eventually discover the new resources and adapt accordingly. On the other hand, Greedy fails to do so and maintains a high distance. 

\noindent\textbf{Robustness against \enquote{greedy} devices.}
We consider a  setting in which 7 devices use Smart EXP3 and 7 devices use Greedy. Figure \ref{figure:static_4_7_22_14users_3networks_dishonestUsers} shows that, on average, those who leverage Smart EXP3 experience a lower distance from the average bit rate available, hence a higher gain, given that it learns continuously and adapts to changes in its environment. On the other hand, Greedy may get stuck in the wrong network even if it experiences a drop in its gain; a device's gain may be different from that of other devices sharing the same network (all devices may not observe an equal share of the network's bandwidth). While simulation shows that 50\% of \enquote{greedy} devices in the environment succeed in performing well, it is not true in a real-world setting (based on results from experiments).
\subsection{Experiments in the wild}\label{section:evaluationExperimentWild}
    We evaluate the performance of Smart EXP3, in comparison to Greedy, through experiments in the wild and observe that Smart EXP3 achieves higher cumulative goodput (faster download). 
The experiments were conducted in a coffee shop, and a selection had to be made between a public WiFi network and a cellular network.
The number of devices and their selection approaches, as well as the bandwidth limit of the networks were unknown. The mobility of devices entering and leaving the service area was not controlled.

Smart EXP3 and Greedy were run sequentially on a laptop, 
equipped with a built-in WiFi interface and connected to the cellular network through a tethered phone. 
The aim was to download a 500MB file \cite{speedTest}, while connecting to the optimal network and optimizing on download time. The load of the public WiFi network and the cellular network, monitored using Wireshark \cite{combs2007wireshark} and by capturing the EcIo values \cite{chakraborty2013coordinating} from the mobile phone, respectively, varied during the experiments. Results from 12 runs of each algorithm show that Smart EXP3 could achieve 1.2x faster download, on average, compared to Greedy. Greedy took 15.67 minutes, on average, to download the file while Smart EXP3 took 12.90 minutes, on average, i.e., Smart EXP3 achieves about 18\% faster download.
\section{Other related work} \label{section:relatedWorks}
Currently, wireless devices have a static preference for WiFi over cellular, and associate with a WiFi Access Point (AP) that has the highest signal strength. However, this is often suboptimal~\cite{biswas2015large}. 
In this section, we discuss state-of-art wireless network selection approaches that have been proposed in the literature, and relevant work done on bandit algorithms.

A significant amount of work leverages the use of multiple wireless networks, such as Multinet \cite{chandra2004multinet}, MPTCP \cite{ford2013tcp}, and Coolspot \cite{pering2006coolspots}. 
However, Coolspot focuses on saving energy by switching between WiFi and Bluetooth. Moreover, identifying the optimal network is crucial for good performance even in Multinet and MPTCP \cite{deng2014wifi}.
A number of centralized approaches \cite{aryafar2017max, bejerano2004fairness, mishra2006client, sui2016characterizing} have been proposed to solve the wireless network selection problem. However, they are not scalable and are limited to managed networks. 
Several distributed solutions have been presented, but they all have some limitations. Some require coordination from APs \cite{kauffmann2007measurement} or cooperation of peers \cite{deng2014all}. Others assume global knowledge \cite{niyato2009dynamics,aryafar2013rat, monsef2015convergence}, or availability of some information \cite{zhu2010network,cheung2017congestion}. In \cite{wu2016traffic}, the problem is formulated as a continuous-time multi-armed bandit, but in a stochastic setting. 

Multi-armed bandit algorithms were initially designed to solve a single-player problem. But the adversarial bandit problem, where an adversary determines the payoff for each arm, can be easily related to a repeated multi-player game \cite{auer2002nonstochastic}. While EXP3 \cite{auer2002nonstochastic} ignores switching cost, the concept of updating in a block manner has been proposed \cite{auer2002finite, du2016learning,chen2011Opportunistic} to take into account switching cost. Multi-armed bandit techniques have also been applied to other resource selection problems, such as channel selection \cite{gai2010learning, tekin2011performance}, selection of the appropriate sensors to query in a sensor network \cite{golovin2010online}, and selection of replica server for content distribution networks \cite{tran2014qoe}. While switching network has a non-trivial cost, the notion of switching cost does not apply to the latter two 
problems. Channel selection approaches do not consider switching cost or require coordination of peers \cite{gai2010learning}. 
%

\section{Conclusion} \label{section:conclusion}

Multi-armed bandit algorithms have impressive theoretical properties that suggest their suitability to solve distributed resource selection problems where coordination among players or support from resource providers is costly or infeasible. Yet, these algorithms do not perform well in practice and, hence, are rarely used. In particular, we have shown that EXP3, one of the leading bandit algorithms, incurs high switching costs, has slow convergence and fails to efficiently adapt to changes in the environment. We have presented Smart EXP3, a novel bandit-style algorithm, and advocate that bandit algorithms can be leveraged to solve resource selection problems by carefully addressing the practical concerns, such as those of EXP3.

We have shown that Smart EXP3 has good theoretical and practical performance. We prove that it has the same convergence and regret properties as EXP3, and bound its expected number of network switches. We evaluate its performance in dynamic wireless network settings, where a mobile device has to select the optimal wireless network for good performance. Empirical results show that it outperforms alternative selection approaches. It stabilizes at the optimal state with reduced switching and without any coordination, gracefully deals with transient behaviors, and achieves fairness among devices. 
This research is a key stepping stone for enhancing connectivity which is of utmost importance for a smart nation.

As future work, we intend to consider other selection criteria, such as application requirements, energy constraints and monetary cost, and evaluate the algorithm for other resource selection problems with non-negligible switching cost, e.g. WiFi channel selection.

\bibliographystyle{plain}
\bibliography{references}

\appendices
\section{Proof of convergence} \label{appendix:convergenceToNE}
    We assume the version of Smart EXP3 without reset and show, following the steps in \cite{tekin2011performance}, that it retains the convergence property of EXP3. 

\setcounter{equation}{0} 

\noindent From algorithm \ref{algorithm:smartEXP3},
\begin{align}
w_i(b + 1) = w_i(b)\ exp\left(\frac{\gamma g_i(b)}{k \overline{p}(b)}\right) \label{proof:convergence_1}
\end{align}

\begin{align}
p_{i}(b) = (1 - \gamma)\frac{w_{i}(b)}{\sum\limits_{j=1}^{k} w_{j}(b)} + \frac{\gamma}{k} \label{proof:convergence_2}
\end{align}

\noindent From (\ref{proof:convergence_2}),
\begin{align}
(1 - \gamma)w_{i}(b) = \sum\limits_{j=1}^{k} w_{j}(b) \left(p_{i}(b) - \frac{\gamma}{k}\right) \label{proof:convergence_3}
\end{align}
\begin{align}
\textnormal{Let } A_i = exp\left(\frac{\gamma g_i(b)}{k \overline{p}(b)}\right) \label{proof:convergence_4}
\end{align}

\noindent We consider the effect of a client's action $i_b$ on the probability of network $i$. We consider both cases when $i_b = i$ and $i_b \neq i$.

\noindent Let's consider case 1: $i_b = i$ \\
\noindent Using (\ref{proof:convergence_1}), (\ref{proof:convergence_2}) and (\ref{proof:convergence_4}),
\begin{align}
p_{i}(b) &= \frac{(1 - \gamma)\ w_i(b)\ A_i}{\sum\limits_{j=1}^{k} w_{j}(b) - w_i(b) + w_i(b)\ A_i} + \frac{\gamma}{k} \nonumber \\
&= \frac{(1 - \gamma)\ w_i(b)\ A_i}{\sum\limits_{j=1}^{k} w_{j}(b) + w_i(b)(A_i - 1)} + \frac{\gamma}{k} \label{proof:convergence_5}
\end{align}

\noindent Substituting (\ref{proof:convergence_3}) in (\ref{proof:convergence_5}),
\begin{align}
p_{i}(b) &= \frac{\sum\limits_{j=1}^{k} w_{j}(b)\ (p_{i}(b) - \frac{\gamma}{k})\ A_i} {\sum\limits_{j=1}^{k} w_{j}(b) + \frac{(A_i - 1)\ \sum\limits_{j=1}^{k} w_{j}(b)\ (p_{i}(b) - \frac{\gamma}{k})}{1 - \gamma}} + \frac{\gamma}{k} \nonumber \\
&= \frac{(p_i(b) - \frac{\gamma}{k})\ A_i}{1 + \frac{(p_i(b) - \frac{\gamma}{k})(A_i - 1)}{1 - \gamma}} + \frac{\gamma}{k} \nonumber
\end{align}

$$\textnormal{Given that }\frac{d(e^{u(x)})}{dx} = e^{u(x)} \frac{d(u(x))}{dx} \textnormal{,}$$
$$\frac{d A_i}{d\gamma} = \frac{g_i(b)}{k\overline{p}(b)}\ A_i$$

\noindent We obtain the continuous time process from the rate of change of $p_i$ with respect to $\gamma$ as $\gamma \rightarrow 0$ and dropping the discrete block index b.

\begin{align}
\dot{p_{i}} &= \lim_{\gamma \rightarrow 0}\ \frac{dp_i}{d\gamma} \nonumber \\
&= \lim_{\gamma \rightarrow 0}\ \frac{d}{d\gamma}\left(\frac{(p_i(b) - \frac{\gamma}{k})\ A_i}{1 + \frac{(p_i(b) - \frac{\gamma}{k})(A_i - 1)}{1 - \gamma}} + \frac{\gamma}{k}\right) \nonumber \\
&= \frac{p_i(b)\ g_i(b)}{k \overline{p}}(1 - p_i(b)) \nonumber \\
&= \frac{p_i\ g_i}{k \overline{p}}(1 - p_i) \label{proof:convergence_6}
\end{align}

\noindent We now consider case 2: $i_b \neq i$ \\
\noindent Using (\ref{proof:convergence_1}), (\ref{proof:convergence_2}) and (\ref{proof:convergence_4}),
\begin{align}
p_{i}(b) &= \frac{(1 - \gamma)\ w_i(b)}{\sum\limits_{j=1}^{k} w_{j}(b) - w_{i_b}(b) + w_{i_b}(b)\ A_{i_b}} + \frac{\gamma}{k} \nonumber \\
&= \frac{(1 - \gamma)\ w_i(b)}{\sum\limits_{j=1}^{k} w_{j}(b) + w_{i_b}(b)(A_{i_b} - 1)} + \frac{\gamma}{k} \label{proof:convergence_7}
\end{align}

\noindent Substituting (\ref{proof:convergence_3}) in (\ref{proof:convergence_7}),
\begin{align}
p_{i}(b) &= \frac{\sum\limits_{j=1}^{k} w_{j}(b)\ (p_{i}(b) - \frac{\gamma}{k})} {\sum\limits_{j=1}^{k} w_{j}(b) + \frac{(A_{i_b} - 1)\ \sum\limits_{j=1}^{k} w_{j}(b)\ (p_{i_b}(b) - \frac{\gamma}{k})}{1 - \gamma}} + \frac{\gamma}{k} \nonumber \\
&= \frac{p_i(b) - \frac{\gamma}{k}}{1 + \frac{(p_{i_b}(b) - \frac{\gamma}{k})(A_{i_b} - 1)}{1 - \gamma}} + \frac{\gamma}{k} \nonumber
\end{align}

\noindent Hence,
\begin{align}
\dot{p_{i}} &= \lim_{\gamma \rightarrow 0}\ \frac{dp_i}{d\gamma} \nonumber \\
&= \lim_{\gamma \rightarrow 0}\ \frac{d}{d\gamma}\left(\frac{p_i(b) - \frac{\gamma}{k}}{1 + \frac{(p_{i_b}(b) - \frac{\gamma}{k})(A_{i_b} - 1)}{1 - \gamma}} + \frac{\gamma}{k}\right) \nonumber \\
&= -\ \frac{p_i(b)\ p_{i_b}(b)\ g_{i_b}(b)}{k \overline{p}} \nonumber \\
&= -\ \frac{p_i\ p_{i_b}\ g_{i_b}}{k \overline{p}} \label{proof:convergence_8}
\end{align}

\noindent Using (\ref{proof:convergence_6}) and (\ref{proof:convergence_8}), we get the expected change in $p_i$.
\begin{align*}
E[\dot{p_i}] &= \overline{p}\ \frac{p_i\ g_i}{k \overline{p}}(1 - p_i) + \sum\limits_{j \in k - \{i\}} \overline{p}(-\ \frac{p_i\ p_j\ g_j}{k \overline{p}}) \nonumber \\
&= \frac{p_i\ g_i}{k}(1 - p_i) + \sum\limits_{j \in k - \{i\}} -\ \frac{p_i\ p_j\ g_j}{k} \nonumber \\
&= \frac{p_i\ g_i}{k}\ \sum\limits_{j \in k - \{i\}} p_j + \sum\limits_{j \in k - \{i\}} -\ \frac{p_i\ p_j\ g_j}{k} \nonumber \\
&= \frac{p_i}{k}\ \sum\limits_{j \in k - \{i\}} (p_j\ g_i - p_j\ g_j) \nonumber \\
&= \frac{p_i}{k}\ \sum\limits_{j \in k - \{i\}} p_j(g_i - g_j) \nonumber \\
\end{align*}

\noindent Taking expectation with respect to other clients' actions
\begin{align*}
\xi_i &= \frac{p_i}{k}\ \sum\limits_{j \in k - \{i\}} p_j(E[g_i] - E[g_j]) \nonumber \\ 
&= \frac{p_i}{k}\ \sum\limits_{j = 1}^{k} p_j(\overline{\overline{g_i}} - \overline{\overline{g_j}}) \nonumber \\ 
\end{align*}

\noindent Given that this replicator dynamics is identical to the ones in \cite{tekin2011performance} and \cite{kleinberg2009multiplicative}, the rest of the proof follows from \cite{kleinberg2009multiplicative}.
\section{Proof of upper bound on number of network switches} \label{appendix:networkSwitchBound}
    \begin{IEEEproof}
As we seek to find an upper bound, we assume that reset periods are of equal lengths, a block length is given by ${(1 + \beta)}^{x} \le \lceil{(1 + \beta)}^{x}\rceil$, and an equal number of time slots are spent in each network. 

\noindent We start by identifying an upper bound on the number of network switches in one reset period. Let $\Delta$ be the number of switch backs (hence, $\Delta$ blocks of length one; aggregate of $\Delta$ time slots), and $f$ be the number of full blocks spent in each network.

\setcounter{equation}{0} 

\noindent Total number of time slots spent in each network
$$ = {(1 + \beta)}^{0} + \cdots + {(1 + \beta)}^{f - 1} $$

\noindent Number of time slots in one reset period = $ \frac{\tau}{t_d} $

\noindent This implies that
$$\left[{(1 + \beta)}^{0} + \cdots + {(1 + \beta)}^{f - 1}\right] * k + k + \Delta \le \frac{\tau}{t_d}$$ 
($k$ time slots for exploration; $\Delta$ time slots for switch backs; there might be a partial block at the end of the reset period, hence $\le$).\\

\noindent Simplifying the equation and solving for $f$, we get
$$ f \le \frac{\log\left(\frac{\beta \tau}{k\ t_d} - \frac{\beta (\Delta + k)}{k} + 1\right)}{\log(1 + \beta)} $$
Since we are looking for an upper bound, we can ignore the positive factor $\frac{\beta}{k}$ (which is $\le 1$) of $\frac{\beta \tau}{k\ t_d}$ and eliminate the positive number $\frac{\beta(\Delta + k)}{k}$ being subtracted from $\frac{\beta \tau}{k\ t_d}$. Hence,
$$ f \le \frac{\log(\frac{\tau}{t_d} + 1)}{\log(1 + \beta)} $$

\centering
\noindent Number of blocks in one period $\le k \cdot f + k + \Delta + 1$ \\
\noindent (The one is to take care of a possible partial block at the end of the reset period).

\centering
\noindent Number of network switches in one reset period 
$$\le \frac{k \cdot \log(\frac{\tau}{t_d} + 1)}{\log(1 + \beta)} + k + \Delta$$
$$\le \frac{3\ k \ \log(\frac{\tau}{t_d} + 1)}{\log(1 + \beta)}$$

\noindent Thus, the expected number of network switches over $T$ is upper bounded as
$$< \frac{T}{\tau}\left(\frac{3\ k \ \log(\frac{\tau}{t_d} + 1)}{\log(1 + \beta)} \right) $$
%
\noindent which concludes the proof.
\end{IEEEproof}

\section{Proof of upper bound on weak regret} \label{appendix:regretBound}
    \begin{IEEEproof}
We assume that reset periods are of equal lengths and 
$B$ is the number of blocks in one reset period.

We start by identifying an upper bound on weak regret for one reset period. We also assume that the algorithm spends the following fractions of time for each type of action: $\omega$ for exploration, $\delta$ for switch back, $\lambda$ for random selection, and $\alpha$ to flip a coin and following which it selects greedily with probability $\frac{1}{2}$ (hence $\omega + \delta + \lambda + \alpha = 1)$. The proof closely relates to that of EXP3 \cite{auer2002nonstochastic} and leverages the following 4 simple facts derived from definitions:

\setcounter{equation}{0} 

$$\hat{g_{i_b}}(b) = \frac{g_{i_b}(b)}{\overline{p}(b)} $$
\begin{align}
\overline{p}(b) &= \frac{\omega}{|explore\_network|} + \delta + \lambda \cdot p_{i_b}(b) + \frac{\alpha}{2} \cdot p_{i_b}(b) + \frac{\alpha}{2} \nonumber \\ 
&> p_{i_b}(b) \cdot \left(\lambda + \frac{\delta}{2}\right) \nonumber \\
& > p_{i_b}(b) \cdot \psi \textnormal{, where $\psi < 1; \psi = (\lambda + \frac{\delta}{2})$} \nonumber
%
%
\end{align}

\noindent Hence,
\begin{align}
\hat{g_{i_b}}(b) &< \frac{g_{i_b}(b)}{\psi \cdot p_{i_b}(b)} 
\label{proof:regret_1}
\end{align}

\noindent Given that $\hat{g_{i}}(b) = 0$ for all actions $i$ except $i_b$,
\begin{align}
\displaystyle\sum_{i=1}^{k} p_i(b)\ \hat{g_i}(b) &= p_{i_b}(b)\ \hat{g_{i_b}}(b) \nonumber \\
&< p_{i_b}(b) \cdot \frac{g_{i_b}(b)}{\psi \cdot p_{i_b}(b)}\ \textnormal{from (\ref{proof:regret_1})} \nonumber\\
&< \frac{g_{i_b}(b)}{\psi} \label{proof:regret_2}
\end{align}

$$g_{i_b}(b) \in [0, l_{i_b}]$$
\begin{align}
\sum_{i=1}^{k} p_i(b)\ {(\hat{g_i}(b))}^2 &< \frac{g_{i_b}(b)}{\psi} \cdot \hat{g_{i_b}}(b)\ \textnormal{from (\ref{proof:regret_2})}\nonumber \\ 
&< \frac{l_{i_b} \cdot \hat{g_{i_b}}(b)}{\psi} \nonumber \\
& < \frac{1}{\psi} \cdot \sum_{i=1}^{k} l_{i}\ \hat{g_i}(b)  \label{proof:regret_3}
\end{align}

\noindent By definition,
\begin{align} 
E[\hat{g_i}(b) | i_1, \cdots, i_b] &= \sum_{i=1}^{k} \hat{g_i}(b)\ \overline{p}(b) \nonumber\\
&= \hat{g_{i_b}}(b)\ \overline{p}(b)\textnormal{, given that }\hat{g}_i(b) = 0\textnormal{ if } i \ne i_b \nonumber \\
&= g_{i_b}(b) \label{proof:regret_4}
\end{align}

\noindent We now proceed with the proof. Let $W_b = w_1(b) + \cdots + w_k(b)$. The proof involves trying to find a bound on the ratio of weights from one round to the next, i.e. $\frac{W_{b+1}}{W_b}$. 

\begin{align}
\frac{W_{b+1}}{W_b} &= \sum_{i=1}^{k} \frac{w_i(b + 1)}{W_b}\textnormal{, given that }W_{b+1} = \sum_{i=1}^{k} w_i(b + 1) \nonumber \\
&= \sum_{i=1}^{k} \frac{w_i(b)}{W_b}\ exp\left(\frac{\gamma\ \hat{g_i}(b)}{k}\right), \label{proof:regret_5} \\
& \textnormal{  using the weight update rule in algorithm \ref{algorithm:smartEXP3}} \nonumber
\end{align}

\noindent Given the probability update rule, we solve for $\displaystyle \frac{w_i(b)}{W_b}$
\begin{align*}
p_i(b) &= (1 - \gamma)\ \frac{w_i(b)}{\displaystyle\sum_{j=1}^{k} w_j(b)} + \frac{\gamma}{k} \\
&= (1 - \gamma)\ \frac{w_i(b)}{W_b} + \frac{\gamma}{k} 
\end{align*}

\begin{align*}
\textnormal{Thus, }\frac{w_i(b)}{W_b} = \frac{p_i(b) - \frac{\gamma}{k}}{1 - \gamma}
\end{align*}

\noindent Combining this with (\ref{proof:regret_5}), we get
\begin{align}
\frac{W_{b+1}}{W_b} &= \sum_{i=1}^{k} \frac{p_i(b) - \frac{\gamma}{k}}{1 - \gamma}\ exp\left(\frac{\gamma\ \hat{g_i}(b)}{k}\right) \label{proof:regret_6}
\end{align}

\noindent From Taylor series,
\begin{align*}
e^x &\le 1 + x + \frac{1}{2}x^2 \\
&\le 1 + x + (e - 2) x^2
\end{align*}

\noindent In our case $\displaystyle x = \frac{\gamma\ \hat{g}_i(b)}{k}$. Combining this with (\ref{proof:regret_6}), we get
\begin{align}
\frac{W_{b+1}}{W_b} &\le \sum_{i=1}^{k} \frac{p_i(b) - \frac{\gamma}{k}}{1 - \gamma}\ \left[1 + \frac{\gamma\ \hat{g}_i(b)}{k} + (e - 2)\left(\frac{\gamma\ \hat{g}_i(b)}{k}\right)^2\right] \nonumber \\
&\le \sum_{i=1}^{k} \frac{p_i(b) - \frac{\gamma}{k}}{1 - \gamma}\ + \frac{\frac{\gamma}{k}}{1 - \gamma}\ \sum_{i=1}^{k} \hat{g}_i(b)\left(p_i(b) - \frac{\gamma}{k}\right) \nonumber \\
&+ \frac{(\frac{\gamma}{k})^2 (e - 2)}{1 - \gamma}\ \sum_{i=1}^{k} (\hat{g}_i(b))^2\left(p_i(b) - \frac{\gamma}{k}\right) \label{proof:regret_7}
\end{align}

\noindent We solve each of the 3 terms in (\ref{proof:regret_7}) individually. Solving the first term, we get
\begin{align}
\sum_{i=1}^{k} \frac{p_i(b) - \frac{\gamma}{k}}{1 - \gamma} &= \frac{1}{1 - \gamma}\ \left(\sum_{i=1}^{k} p_i(b) - \sum_{i=1}^{k} \frac{\gamma}{k}\right) \nonumber \\
&= \frac{1}{1 - \gamma} (1 - \gamma) \nonumber \\
& = 1   \label{proof:regret_8}
\end{align}

\noindent We now solve the second therm. As we seek to find an upper bound, we can eliminate the positive number $\frac{\gamma}{k}$ being subtracted from $p_i(b)$.
\begin{align}
\frac{\frac{\gamma}{k}}{1 - \gamma}\ \sum_{i=1}^{k} \hat{g}_i(b)(p_i(b) - \frac{\gamma}{k}) &< \frac{\frac{\gamma}{k}}{1 - \gamma}\ \sum_{i=1}^{k} \hat{g}_i(b)\ p_i(b) \nonumber \\
&< \frac{\frac{\gamma}{k}}{1 - \gamma} \cdot \frac{g_{i_b}(b)}{\psi} \textnormal{ from (\ref{proof:regret_2})} \nonumber \\
&< \frac{\left(\frac{\gamma}{k}\right) g_{i_b}(b)}{\psi(1 - \gamma)} 
\label{proof:regret_9} \nonumber \\
\end{align}

\noindent We solve the third term, again ignoring the positive number $\frac{\gamma}{k}$ being subtracted from $p_i(b)$.
\begin{align}
\frac{(\frac{\gamma}{k})^2 (e - 2)}{1 - \gamma}\ &\sum_{i=1}^{k} \left(\hat{g}_i(b)\right)^2\left(p_i(b) - \frac{\gamma}{k}\right) \nonumber \\
&< \frac{(\frac{\gamma}{k})^2 (e - 2)}{1 - \gamma}\ \sum_{i=1}^{k} (\hat{g}_i(b))^2\ p_i(b) \nonumber \\
&< \frac{(\frac{\gamma}{k})^2 (e - 2)}{\psi (1 - \gamma)} \sum_{i=1}^{k} l_i\ \hat{g_i}(b) \textnormal{ from (\ref{proof:regret_3})}
\label{proof:regret_10}
\end{align}

\noindent Combining (\ref{proof:regret_8}), (\ref{proof:regret_9}) and (\ref{proof:regret_10}) in (\ref{proof:regret_7}), we get
\begin{align}
\frac{W_{b+1}}{W_b} &\le 1 + \frac{\left(\frac{\gamma}{k}\right) g_{i_b}(b)}{\psi(1 - \gamma)} + \frac{(\frac{\gamma}{k})^2 (e - 2)}{\psi (1 - \gamma)} \sum_{i=1}^{k} l_i\ \hat{g_i}(b) \nonumber
\end{align}

\noindent Taking logarithms on both sides,
\begin{align}
\ln \frac{W_{b+1}}{W_b} &\le \ln\left(1 + \frac{\left(\frac{\gamma}{k}\right) g_{i_b}(b)}{\psi(1 - \gamma)} + \frac{(\frac{\gamma}{k})^2 (e - 2)}{\psi (1 - \gamma)} \sum_{i=1}^{k} l_i\ \hat{g_i}(b)\right) \label{proof:regret_11}
\end{align}

$$1 + a \le e^a \textnormal{, when } a > 1$$
In our case,
$\displaystyle a = \frac{\left(\frac{\gamma}{k}\right) g_{i_b}(b)}{\psi(1 - \gamma)} + \frac{(\frac{\gamma}{k})^2 (e - 2)}{\psi (1 - \gamma)} \sum_{i=1}^{k} l_i\ \hat{g_i}(b) $ \\ 
\noindent Hence, from (\ref{proof:regret_11})
\begin{align}
\ln\ W_{b+1} - \ln\ W_b &\le \frac{\left(\frac{\gamma}{k}\right) g_{i_b}(b)}{\psi(1 - \gamma)} + \frac{(\frac{\gamma}{k})^2 (e - 2)}{\psi (1 - \gamma)} \sum_{i=1}^{k} l_i\ \hat{g_i}(b) \nonumber 
\end{align}

\noindent Summing over b
\begin{align}
\sum_{b=1}^B\ (\ln\ W_{b+1} - \ln\ W_b) &\le \frac{\frac{\gamma}{k}}{\psi(1 - \gamma)} \sum_{b=1}^B g_{i_b}(b) \nonumber \\
&+ \frac{(\frac{\gamma^2}{k}) (e - 2)}{\psi (1 - \gamma)} \sum_{b=1}^B \sum_{i=1}^{k} l_i\ \hat{g_i}(b)
\label{proof:regret_12}
\end{align}

\begin{align}
W_{B+1} &\ge w_j(B+1) \nonumber \\
w_j(B&+1) = w_j(B)\ exp\left(\frac{\gamma\ \hat{g_j}(B)}{k}\right) \nonumber \\
&= w_j(B-1)\ exp\left(\frac{\gamma\ \hat{g_j}(B-1)}{k}\right)\ exp\left(\frac{\gamma\ \hat{g_j}(B)}{k}\right) \nonumber \\
&=\prod_{b=1}^B\ exp\left(\frac{\gamma\ \hat{g_j}(b)}{k}\right) \nonumber \\
&=exp\left(\frac{\gamma}{k}\ \sum_{b=1}^B\ \hat{g_j}(b)\right) \nonumber \\
W_{B+1} &\ge exp\left(\frac{\gamma}{k}\ \sum_{b=1}^B\ \hat{g_j}(b)\right) \nonumber
\end{align}

\noindent Taking logarithms on both sides
\begin{align}
\ln\ W_{B+1} \ge \frac{\gamma}{k}\ \sum_{b=1}^B\ \hat{g_j}(b) \label{proof:regret_13}
\end{align}

\noindent Simplifying the left-hand side of (\ref{proof:regret_12}), which is a telescoping sum, and combining (\ref{proof:regret_13}), we have
\begin{align}
\sum_{b=1}^B\ (\ln\ W_{b+1} - \ln\ W_b) &= \ln\ W_{B+1} - \ln\ W_1 \nonumber \\
&\ge \frac{\gamma}{k}\ \sum_{b=1}^B\ \hat{g_j}(b) - \ln\ k \label{proof:regret_14}
\end{align}

\noindent $g_{i_b}(b)$ refers to gain from Smart EXP3 in block b. Summing $g_{i_b}(b)$ over B gives the total gain of the algorithm, $G_{SmartEXP3}(B)$. We combine this and (\ref{proof:regret_14}) with (\ref{proof:regret_12}).
\begin{align}
\frac{\gamma}{k}\ \sum_{b=1}^B\ \hat{g_j}(b) - \ln\ k &\le \frac{\frac{\gamma}{k}}{\psi(1 - \gamma)} G_{SmartEXP3}(B) \nonumber \\
&+ \frac{(\frac{\gamma^2}{k}) (e - 2)}{\psi(1 - \gamma)} \sum_{b=1}^B \sum_{i=1}^{k} l_i\ \hat{g_i}(b) \nonumber
\end{align}

\noindent Multiplying both sides by $\displaystyle \frac{\psi(1 - \gamma)}{\frac{\gamma}{k}}$ and simultaneously solving for $G_{SmartEXP3}(B)$,
\begin{align}
G_{SmartEXP3}(B) &\ge \psi(1 - \gamma) \sum_{b=1}^B\ \hat{g_j}(b) - \frac{\psi(1 - \gamma)}{\frac{\gamma}{k}}\ \ln\ k \nonumber \\
&- \frac{\gamma}{k}\ (e - 2)\ \sum_{b=1}^B\ \sum_{i=1}^{k} l_i\ \hat{g_i}(b) \nonumber
\end{align}

\noindent Taking expectation on both sides
\begin{align}
&E\left[G_{SmartEXP3}(B)\right] \ge \nonumber \\
& \psi(1 - \gamma) \sum_{b=1}^B\ E\left[\hat{g_j}(b)\right] - \frac{\psi(1 - \gamma)}{\frac{\gamma}{k}}\ \ln\ k \nonumber \\
&- \frac{\gamma}{k} \cdot (e - 2) \cdot l\ \sum_{b=1}^B\ \sum_{i=1}^{k} E\left[\hat{g_i}(b)\right] \label{proof:regret_15}
\end{align}
where $l$ is the largest block length.

\noindent Using (\ref{proof:regret_4}), $\sum_{b=1}^B\ E[\hat{g_j}(b)] = G_{max}(B)$ if we pick the best action $j$ \\
\noindent Combining this with (\ref{proof:regret_15})
\begin{align}
&E\left[G_{SmartEXP3}(B)\right] \ge \nonumber \\
& \psi(1 - \gamma) G_{max}(B) - \frac{\psi(1 - \gamma)}{\frac{\gamma}{k}}\ \ln\ k \nonumber \\
&- \frac{\gamma}{k} \cdot (e - 2) \cdot l\ \sum_{b=1}^B\ \sum_{i=1}^{k} E\left[\hat{g_i}(b)\right] 
\label{proof:regret_16}
\end{align}

\begin{align*}
\sum_{b=1}^B\ \sum_{i=1}^{k} \ E[\hat{g_i}(b)] &= \sum_{b=1}^B\ \sum_{i=1}^{k} \ g_i(b) \textnormal{ from (\ref{proof:regret_4})} \nonumber \\
&=\sum_{i=1}^{k}\ \sum_{b=1}^B \ g_i(b) \textnormal{ switching b and i sums}
\end{align*}

$$\sum_{b=1}^B \ g_i(b) \le G_{max}(B) \textnormal{ if i is fixed; i is at most the best action}$$
$$\sum_{i=1}^{k}\ \sum_{b=1}^B \ g_i(b) \le k\ G_{max}(B)$$
\noindent Combining this with (\ref{proof:regret_16}),
\begin{align*}
E\left[G_{SmartEXP3}(B)\right] &\ge \psi(1 - \gamma) G_{max}(B) - \frac{\psi(1 - \gamma)}{\frac{\gamma}{k}}\ \ln\ k \nonumber \\
&- \gamma \cdot (e - 2) \cdot l\ G_{max}
(B)\nonumber \\
&\ge (\psi(1 - \gamma) - \gamma \ l\ (e - 2))\ G_{max}(B) \nonumber \\
&- \frac{\psi(1 - \gamma)}{\frac{\gamma}{k}}\ \ln\ k \nonumber
\end{align*}

\noindent Subtracting $G_{max}$ from both sides,
\begin{align*}
E[G_{SmartEXP3}(B)] &- G_{max}(B) \nonumber \\
& \ge (\psi(1 - \gamma) - \gamma \ l\ (e - 2) - 1)\ G_{max}(B) \nonumber \\
&- \frac{\psi(1 - \gamma)}{\frac{\gamma}{k}}\ \ln\ k \nonumber
\end{align*}

\noindent Flipping the inequality,
\begin{align*}
G_{max}(B) &- E[G_{SmartEXP3}(B)] \nonumber \\
& \le (1 + \gamma \ l\ (e - 2) - \psi(1 - \gamma))\ G_{max}(B) \nonumber \\
&+ \frac{\psi(1 - \gamma)}{\frac{\gamma}{k}}\ \ln\ k \nonumber \\
& \le (1 + \gamma \ l\ (e - 2))\ G_{max}(B) + \frac{k}{\gamma}\ \ln\ k \nonumber
\end{align*}

\begin{align*}
G_{max}(\tau) &- E[G_{SmartEXP3}(\tau)] \nonumber \\
& \le (1 + \gamma \ l\ (e - 2))\ G_{max}(\tau) + \frac{k\ ln\ k}{\gamma} \nonumber
\end{align*}

\noindent Given that there are $\frac{T}{\tau}$ reset periods,
\begin{align}
&G_{max}(T) - E[G_{SmartEXP3}(T)] \nonumber \\
&\le \frac{T}{\tau}\left( (1 + \gamma \ l\ (e - 2))\ G_{max}(\tau) + \frac{k\ ln\ k}{\gamma}\right) \nonumber 
\end{align}

\noindent Since gain ignores switching cost, 
\begin{align}
E[R(T)] =\ &(G_{max}(T) - E[G_{SmartEXP3}(T)]) \cdot t_d \nonumber \\
&+ E[S(T)] \cdot \mu_d \cdot \mu_g \nonumber \\
\end{align}

\begin{align}
E[R(T)] &\le \frac{T \cdot t_d}{\tau}\left((1 + \gamma \ l\ (e - 2))\ G_{max}(\tau) + \frac{k\ ln\ k}{\gamma}\right) \nonumber \\
&+ \frac{T \cdot \mu_d \cdot \mu_g}{\tau}\left(\frac{3\ k \ \log(\frac{\tau}{t_d} + 1)}{\log(1 + \beta)}\right)  \nonumber \\
\end{align}
which concludes the proof.
\end{IEEEproof}
\label{appendix:algorithms}

\end{document}